\documentstyle[12pt]{article}
\begin{document}

\def\p{\partial}
\def\be{\begin{equation}}
\def\bea{\begin{eqnarray}}
\def\ee{\end{equation}}
\def\eea{\end{eqnarray}}
\def\bearst{\begin{eqnarray*}}
\def\eearst{\end{eqnarray*}}
\def\dbar{\bar \partial}
\def\nn{\nonumber}
\def\ll{\lambda}
\def\l{\label}
\def\D{\Delta}
\def\o{\over}
\def\E{{\rm e}}
\def\peleven{\parbox{14cm}}
\def\peffec{\peight{\bearst\eearst}\hfill\peleven}
\def\pspace{\peight{\bearst\eearst}\hfill}
\def\ptwelve{\parbox{15cm}}
\def\peight{\parbox{8mm}}
        \def\sl{\hbox{{\rm sl}}}
     \def\half{{1 \over 2}}
     \def\Uq#1{{\rm U}_q \left( #1 \right) }
     \def\Um#1{{\rm U}_{\mu} \left( #1 \right) }
     \def\U#1{{\rm U} \left( #1 \right) }
     \def\D{{\Delta }}
     \def\P{{\Phi}}
     \def\p{{\phi}}
     \def\si{{\psi}}
     \def\d{{\partial}}
     \def\R{{\bf R}}
     \def\s{\sum}
     \def\pr{\prod}
     \def\a{\alpha}
     \def\eps{\epsilon}
     \def\HG{\H_{"G"}}
     \def\Ts{\tilde{\psi}}
     \def\Si{\Psi}
     \def\Raw{\Rightarrow}
     \def\raw{\rightarrow}
     \def\da{\dag}
     \def\lm{\lambda}
     \def\sq{\sqrt}
     \def\D{\Delta}
     \def\b{\beta}
     \def\i{\imath}
     \def\im{\rm {\it Im}}
     \def\re{\rm {\it Re}}
     \def\ah{\hat{A}}
     \def\bh{\hat{B}}
     \def\ch{\hat{C}}
     \def\eh{\hat{D}}
     \def\c{\cdot}
     \def\cs{\cdots}
     \def\e{\epsilon}
     \def\g{\gamma}
\def\CN{{\cal N}}
\def\CM{{\cal M}}
\def\CA{{\cal A}}
\def\CB{{\cal B}}
\def\CD{{\cal D}}
\def\CO{{\cal O}}
\def\CD{{\cal D}}
\def\W{{\cal W}}
\def\lk{\left [}
\def\rk{\right ]}
\def\kt#1{\mid{{#1}}>}
\def\br#1{<{#1}\mid}
\def\z#1{z_{#1}}
\def\de#1{\Delta_{#1}}
\def\ds{\Delta_\Psi}
\def\df{\Delta_\Phi}
\def\da{\Delta_A}
\def\db{\Delta_B}
\def\dc{\Delta_p}
\def\I{\rm {I\kern-.3em I}}
\def\C{\rm {I\kern-.520em C}}
\def\R{\rm {I\kern-.3em R}}
\def\CZ{\rm {Z\kern-.4em Z}}
\def\str{$sl(2,\C)$}
 \def\ps#1{\psi({#1})}
\def\unit{\rm {1\kern-.4em 1}}
\def\f{\frac}
    \def\ff{\f{1}{2}}
    \def\pp{<\Psi(\z1)\Psi(\z2)>}
\def\abf{<A(z_1)B(z_2)\Phi(z_3)>}
\def\abs{<A(z_1)B(z_2)\Psi(z_3)>}
\def\ab#1{\vert #1 \vert}
\def\kp{\kt{\de{p}}}
\def\kpp{\kt{{\D'}_p}}
\def\kk{{\bf k}}
\def\E{{\rm e}}
\def\peleven{\parbox{11cm}}
\def\peffec{\peight{\bearst\eearst}\hfill\peleven}
\def\pspace{\peight{\bearst\eearst}\hfill}
\def\ptwelve{\parbox{12cm}}
\def\peight{\parbox{8mm}}
\title{ Disordered Systems and
Logarithmic Conformal Field Theory
}
\author{M. Reza Rahimi Tabar \footnote{
on leave, Dept. of Physics , Iran  University of Science and
Technology, Narmak, Tehran 16844, Iran.}\\
Department of Physics, Sharif University of Technology,\\
P.O.Box 11365--9161, Tehran, Iran\\
and\\
 CNRS UMR 6529, Observatoire de la C$\hat o$te d'Azur,\\
BP 4229, 06304 Nice Cedex 4, France}\maketitle

\vskip 1cm

\begin{center}
{\small \tt Lectures at the International Summer School on
 Logarithmic Conformal Field Theory and Its Applications,
Sept. 2001 Tehran}
\end{center}

\begin{abstract}

We review a recent development in theoretical
understanding of the quenched averaged correlation functions
of disordered systems and the logarithmic conformal field theory (LCFT)
 in d-dimensions.
The logarithmic conformal
field theory is the generalization of the conformal field theory
 when the dilatation operator
is not diagonal and has the Jordan form. It is discussed that at
the random fixed point the disordered systems such as random-bond
Ising model, Polymer chain, etc. are described by LCFT and their
correlation functions have logarithmic singularities. As an
example we discuss in detail the application of LCFT to the
problem of random-bond Ising model in  $ 2 \leq  d  \leq 4$.\\
{{ PACS, 05.70.jk;11.25.Hf;64.60.Ak}}
\end{abstract}

\maketitle
\newpage

\section{Introduction}

Random systems represent the spatial inhomogeneity where scale
invariance is only preserved on average but not for specific
disorder realization. The understanding of the role played by
quenched impurities of the nature of phase transition is one of
the significant subjects in statistical physics and has attracted
a great deal of attention [1]. According to the Harris criterion
[2], quenched randomness is a relevant perturbation at the
second-order critical point for systems of dimension $d$, when its
specific heat exponent $\alpha$, of the pure system is positive.
Concerning the effect of randomness on the correlation functions,
it is known that the presence of randomness induces a logarithmic
factor to the correlation functions of pure system [3-4].
Theoretical treatment of the quenched disordered systems is a
non-trivial task in view of the fact that, one has to average the
logarithm of the partition function over various realization of
the disorder in the statistical ensemble and therefore find
physical quantities [1]. There are two standard methods to perform
this averaging, the supersymmetry (SUSY) approach, and the
well-known replica approach. Recently using the replica approach
it has been shown by Cardy [5], that the logarithmic factor
multiplying power law behavior are to
 be expected in the scaling behavior near non-mean field critical points (see also [54]).
It is shown also that the results are valid for systems with
short-range interactions and in an arbitrary number of dimensions.
He concludes that in the
 limit of $n \rightarrow 0$ of replicas the theory possess of a set of
 fields which
 are degenerate (they have the same scaling dimensions)
  and finds a pair of fields which form
 a Jordan cell structure for dilatation operator and derives
 logarithmic operator in such disordered systems.
It is proved that the
 quenched disordered theory with $Z=1$ can be described by logarithmic
 conformal field theory as well. The logarithmic conformal field theories
 (LCFT) [6-7]
 are extensions of
 conventional conformal field theories [8-10], which have emerged in
 recent years
  in a number of interesting physical problems of
  WZNW models [11-15], supergroups and super-symmetric field theories [16-22]
  Haldane-Rezzayi state in the fractional quantum Hall effect [23-27],
  multi-fractality [28], two-dimensional turbulence [29-31], gravitaitionally dressed
  theories [32], Polymer and abelian sandpiles [33-35, 5],
   String theory and D-brane recoil [36-44],
  Ads/CFT correspondence [45-52],
   Seiberg-Witten solution to SUSY Yang-Mills theory
  [53], disordered systems [54-63]. Also the material such as
  Null vectors, Characters, partition functions, fusion rules, Modular
  Invariance, C-theorem, LCFT`s with boundary and operator product
  expansions have been discussed in
  [64-84].

The LCFT are characterized by the fact that their dilatation operator $L_0$
are not diagonalized and admit a Jordan cell structure.
The non-trivial mixing between these operators leads to logarithmic
singularities in their correlation functions. It has been shown [6]
that the correlator
of two fields in such field theories, has a logarithmic singularity
as follows,

\be
<\psi(r) \psi(r^{'})>\sim {|r-r^{'}|}^{-2 \Delta_{\psi}} \log|r-r^{'}| + \ldots
\ee

 In this article we review the conformal field
 theory (CFT) and
 logarithmic conformal field theory (LCFT), which have
  appeared in the last
decade as a powerful tool for the description of the correlation
functions of second-order phase transition of pure and disordered
critical systems near their fixed points. In section 2 and 3 we
present a brief and self-content review of the CFT and LCFT and
their basic tools in d-dimensions. In section 4 using the replica
method we show that the disordered systems near their fixed point
can be described by LCFT. As an example we discuss in details the
correlation functions of the random-bond Ising model and its
connection to LCFT. We give the explicit expression of the various
types of quenched averaged 2, 3 and 4-point
 correlation functions
 of the local energy density. We also show that the ratios of these
 correlation functions
 to the connected ones have specific universal asymptotic and write
 down these universal functions explicitly.


\begin{center}
\section{ \LARGE  - Conformal Field Theory}
\end{center}

In the following sub-sections we introduce the necessary
techniques and the basic definition such as
 conformal transformation, conformal
group, its representation, correlation functions, Ward-identities,
Virasoro algebra and its representation in 2d conformal field theories, etc.

\begin{center} {\large $\bullet$ Conformal Transformation $\&$
Conformal Group}
\end{center}
Let us start with definition of conformal transformation.

  Definition-1 : A transformation of coordinates ${\bf x'}
\rightarrow {\bf x} $
is called conformal if it leaves the d-dimensional metric $g_{\mu,\nu}$
unchanged up to a scalar factor $\Lambda(x)$, i.e.
{
\be
 g_{\mu \nu}'(x) = \Lambda(x) g_{\mu \nu}(x)
\ee
}
{ consider an infinitesimal transformation }
{
\be
{ x^{\mu} \rightarrow x'^{\mu } = x^{\mu} + \epsilon^{\mu}(x)}
\ee
}
{ to order $(\epsilon)$:}
{
\be
{   g_{\mu \nu} \rightarrow g_{\mu \nu}- ( \epsilon_{\mu;\nu} +
\epsilon_{\nu;\mu}) }
\ee
}
{ where $\epsilon_{\alpha;\beta}$ is the covariant derivative of
$\epsilon_\alpha$.
 The condition that the transformation be conformal gives: }

{
\be
{ \epsilon_{\mu;\nu} + \epsilon_{\nu;\mu} = f(x) g_{\mu \nu}}
\ee
}

{ where}

\begin{center}
{ $f(x) = \Lambda(x) -1 $}
\end{center}

{ Now we can eliminate $f(x)$. Let us suppose that
$g_{\mu \nu} = \eta_{\mu \nu} = Diag(1,-1,\cdots)$. Contract eq. (5) with
$\eta_{\mu \nu}$ to get:}

{
\be
{ f(x) = \frac{2}{d} ( \partial_\nu \epsilon_\nu) = \frac{2}{d} (
\partial \cdot {\bf \epsilon}) }
\ee
}
{  This gives us the first relation between $ f(x)$ and ${\bf \epsilon}$.
Differentiate eq. (5) w.r.t. $x^{\alpha}$ and
after reordering of indices:
}
{
\be
{ 2\partial_{\mu} \partial_{\nu} \epsilon_\alpha = \eta_{\mu \alpha} \partial_{\nu} f}
+ \eta_{\nu \alpha} \partial_{\mu} f - \eta_{\mu \nu} \partial_{\alpha} f
\ee
}
{  Now contract (7) with $\eta^{\mu \nu}$ gives: }
{
\be
{ 2 \partial^2 \epsilon_\mu = (2-d) \partial_\mu f }
\ee
}
{  To find final equation for $f$ apply $\partial_\mu$ to eq.(8)
and $\partial^2$ to eq.(5) and find: }
{
\be
{ (2-d) \partial_\mu \partial_\nu f = \eta_{\mu \nu} \partial^2 f}
\ee
}
{ now contract above eq. with $\eta^{\mu \nu}$ (using $\eta^{\mu \nu}
\eta_{\mu \nu} = d $), therefore we find that the $f$ satisfies the
following equation :
}
{
\be
{ 2(1-d) \partial^2 f = 0 }
\ee
}

\begin{center}
{ \large $\bullet$  Conformal Group with $ {\small d} > 2$}
\end{center}

For $d>2$ eq. (10) reduces to following simple equation,
{
\be
{ \partial^2 f = 0 \hskip .5cm \Rightarrow \hskip .5 cm f(x) = A + B_\mu x^{\mu}}
\ee
}
{ Using the relation
$f(x) = \frac{2}{d} \partial_\alpha \epsilon^{\alpha}$:}

{
\be
 \epsilon_\mu = a_\mu + b_{\mu \nu} x^\nu + c_{\mu \nu \alpha}
x^\nu x^\alpha
\ee
}

{ The case $ b \equiv c \equiv 0 \hskip 0.5 cm \Rightarrow$ infinitesimal
$\hskip 0.4 cm$ translation.
 }{ Now suppose $a \equiv c \equiv 0$. Substitute $ \epsilon_\mu = b_{\mu \nu}
x^\nu$ in eq. (5) gives:}

{
\be
 b_{\mu \nu} + b_{\nu \mu} = \frac{2}{d} b_{\alpha} ^{\alpha}
 \eta_{\mu \nu}
\ee
}

{ Therefore $b_{\mu \nu}$ has two part which are antisymmetric
and proportional to $\eta_{\mu \nu}$, i.e.
}

{
\be
 b_{\mu \nu} = \lambda \eta_{\mu \nu} + m_{\mu \nu} \hskip 1 cm
m_{\mu \nu } = - m_{\nu \mu}
\ee
}
 where $\lambda$ is scaling factor. The term which is proportional
to $\eta_{\mu \nu}$ is scaling transformation and $m_{\mu \nu}$ part
is an infinitesimal rotation. To understand the meaning of
the part $c_{\mu \nu \alpha} x^\nu x^\alpha$ we start with eq.(9) for $d > 2$:
\be
 (2-d) \partial_\mu \partial_\nu f = \eta_{\mu \nu} \partial^2 f = 0
\Rightarrow \partial_\mu \partial_\nu f = 0
\ee
 or:
{
\be
{ \partial_\mu \partial_\nu \partial \cdot {\bf \epsilon} = 0}
\ee
}
{
\be
{ \partial_\nu \partial \cdot \epsilon = - 2 b_\nu \Rightarrow
{\bf \epsilon^{\mu}}
= b^{\mu} {\bf x \cdot x} - 2 x^\mu {\bf b\cdot x}}
\ee

{  This is known as special conformal transformation which has
combination of inversion and translation. For finite transformation:
}
\bea
{ x'} ^{\mu}  && = x^ \mu + a^ \mu \cr \nonumber \\
       { x'} ^{\mu}  && = \lambda x^\mu \cr \nonumber \\
       { x'} ^{\mu} && = m_{\nu} ^{\mu} x^\nu  \cr \nonumber \\
       { x'} ^{\mu}  && = \frac{x^\mu - b^\mu x^2}{1-2{\bf b \cdot x} + b^2 x^2}
\eea
\vskip 1cm
{ Generators of conformal  group are: (for $d>2$) }
\vskip 1cm
{
\bea
p_\mu = && - i \partial_\mu \cr \nonumber \\
D=&&  - i x^\mu \partial_\mu \cr \nonumber \\
L_{\mu \nu} =&& i ( x_\mu \partial_\nu - x_\nu \partial_\mu) \cr \nonumber \\
K_\mu =&& - i ( 2 x_\mu x^\nu \partial_\nu - x^2 \partial_\mu)
\eea
}

{and the algebra is}

{
\bea
\left[ D, P_\mu \right] =&& i P_\mu \cr \nonumber \\
\left[ D,K_\mu \right] = && - i K_\mu \cr \nonumber \\
\left[K_\mu, P_\nu \right] = && 2 i ( \eta_{\mu \nu} D - L_{\mu \nu}) \cr \nonumber \\
\left[K_\rho, L_{\mu \nu} \right] =&& i (\eta_{\rho \mu} K_\nu - \eta_{\rho \nu} K_\mu)
\cr \nonumber \\
\left[ P_\rho, L_{\mu \nu} \right] =&& i ( \eta_{\rho \mu} P_\nu - \eta_{\rho \nu} P_\mu )
\cr \nonumber \\
\left[ L_{\mu \nu}, L_{\rho \sigma} \right] =&&
 i (  \eta_{\nu \rho} L_{\mu \sigma}
+ \eta_{\mu \sigma} L_{\nu \rho}
-  \eta_{\mu \rho} L_{\nu \sigma}
-  \eta_{\nu \sigma} L_{\mu \rho})
\eea
}
{  Define new generators as:}
{
\bea
J_{\mu \nu} = && L_{\mu \nu} \cr \nonumber \\
J_{-1, \mu} = && \frac{1}{2} ( P_\mu - K_{\mu}) \cr \nonumber \\
J_{-1,0} = && D  \cr \nonumber \\
J_{0,\mu} = && \frac{1}{2} ( P_\mu + K_{\mu})
\eea
}
{ Note that $J_{a,b} = - J_{b,a} $ and $a,b \in \{ -1,0,1,\cdots,d \}$.
$J_{a,b}$ satisfy the $SO(d+1,1)$ algebra. Number of its parameters
is $\frac{1}{2} (d+2)(d+1)$. $J_{a,b}$ satisfy: }
{
\be
\left[ J_{a,b}, J_{c, d} \right] = i (  \eta_{a d} J_{b c}
+ \eta_{b c } J_{a d}
-  \eta_{a c} J_{b d }
-  \eta_{b d } J_{a c} )
\ee
}
\begin{center}
{ $\bullet$ Representation of Conformal Group}
\end{center}
{
Consider infinitesimal transformation with parameters $\omega^g$.
 We would like to find the representation of $T_g$ so that:}
{
\be
{\Phi'(x') = ( 1 - i \omega^g T_g) \Phi(x)}
\ee
}
{ Now define:}
{
\be
{L_{\mu \nu} \Phi(0) = S_{\mu \nu} \Phi(0)}
\ee
}
{ Near the origin:}
{
\bea
&& \exp( i x^\alpha P_\alpha) L_{\mu \nu}   \exp( -  i x^\alpha P_\alpha)
\cr \nonumber \\
&& = S_{\mu \nu} - x_\mu P_\nu + x_\nu P_\mu
\eea
}
{ using:}
{
\bea
P_\mu \Phi(x) = - i \partial_\mu \Phi(x)
\eea
}
one finds,
{
\bea
L_{\mu \nu} \Phi(x) &=& i ( x_\mu \partial_\nu - x_\nu \partial_\mu) \Phi(x) +
S_{\mu \nu} \Phi(x)
\eea
}
{  Similarly define the effect of $D$ and $K_\mu$
on the origin as $ \Delta$ and
$k_\mu$ respectively, leads to:}
{
\bea
&&K_\mu \Phi(x) = \cr \nonumber \\
&& ( k_\mu + 2 x_\mu \Delta - x^\nu S_{\mu \nu} -
 2 i x_\mu x^\nu \partial_\nu + i x^2 \partial_\mu) \Phi(x) \cr \nonumber \\
&&D \Phi(x) = ( \Delta - i x^\nu \partial_\nu) \Phi(x)
\eea
}
Now for finite transformation we can define the
quasi-primary field as following.

{  Definition 2 : The field $\Phi(x)$ which under conformal
transformation transforms as:} { \be \Phi(x) \rightarrow \Phi'(x')
= \Omega(x)^{\Delta_i/2} \Phi(x) \ee } { is called {\it
quasi-primary filed} with scaling dimension $\Delta_i/2$. The
relation between $\Omega(x)$ and the Jacobian is, $ |\frac{
\partial x'}{\partial x}| = \Omega^{-d/2}$.}
\\
\begin{center}
{\large $\bullet$ Correlation Functions}
\end{center}
\vskip .5 cm
 Correlation functions of the conformal field theory should transform as
following:
{
\bea
&&< \Phi'(x_1 ')  \cdots \Phi'(x_N ') > = \cr \nonumber \\
&&\Omega(x_1)^{\Delta_1/2} \cdots
\Omega(x_N)^{\Delta_N/2}
 < \Phi(x_1 ) \cdots \Phi(x_N) >
\eea
}
 The conformal structure implies strong constraints on the correlation
functions of the theory. These constrains can be found by using the
infinitesimal transformation and the above transformation.

{One can show that the two, three and four point correlation functions
of $\Phi(x)$ are given by:}
{
\bea
&&< \Phi_1 (x_1) \Phi_2 (x_2) > \cr \nonumber \\
&& = \cases{ \frac{c_{1,2}}{ |x_1 - x_2| ^{ 2 \Delta_1}},
&$\Delta_1 = \Delta_2$  \cr
       0,& $ \Delta_1  \neq  \Delta_2$\cr}
\eea
}

\bea
&& < \Phi_1 (x_1) \Phi_2 (x_2) \Phi_3 (x_3) > = \cr \nonumber \\
&&\frac{c_{1,2,3}}{
x_{1,2} ^{\Delta_1 + \Delta_2 - \Delta_3}
x_{2,3} ^{\Delta_2 + \Delta_3 - \Delta_1}
x_{1,3} ^{\Delta_1 + \Delta_3 - \Delta_2}}
\eea

{
\bea
&&< \Phi_1 (x_1) \Phi_2 (x_2) \Phi_3 (x_3) \Phi_4 (x_4) > = \cr \nonumber \\
&& f( \frac{ x_{1,2}x_{3,4}}{x_{1,3} x_{2,4}},
 \frac{ x_{1,2}x_{3,4}}{x_{2,3} x_{1,4}}, \cdots) \Pi_{i<j}
 ^4 x_{i,j}^ { \Delta/3 - \Delta_i - \Delta_j}
\eea
}
{  where $x_{i,j} = | x_i - x_j|$ and $\Delta= \Sigma_{i=1} ^ 4
\Delta_i$.}
\begin{center}
{\large $\bullet$ The Ward Identities}
\end{center}
{ Consider an infinitesimal transformation as:} { \be \Phi'(x') =
( 1 - i \eta^a(x) G_a) \Phi(x) \ee } { where $G_a$`s are the
generators of group and $\eta^{a}(x)$`s are infinitesimal
functions. Under this transformation action changes as: } { \be
\delta S = - \int d^d x \partial_\mu ( j_a ^\mu \eta^a(x)) \ee } {
where $j_a ^\mu $ is conserved current corresponding
 to the transformation (34). On the other hand we can find the change of
 N-point correlation functions under this transformation. Defining
  $\Phi(N) = \Phi(x_1) \cdots \Phi(x_N) $, it can be shown
   that
 to order $\eta(x)$:}
{
\be
 < \delta \Phi(N) > = - \int d^d x  \partial_\mu < j_a ^\mu \Phi(N) > \eta^a (x).
\ee
}
{  Also using explicit expression of transformation (i.e .eq. (34)),
: }
{
\bea
&&\delta \Phi(N) = -i \sum_{i=1} ^{N} (\Phi(x_1) \cdots G_a  \Phi(x_i)
\cdots \Phi(x_N)) \eta^a(x_i) \cr \nonumber \\
&=&
-i \int d^dx \eta^a(x) \sum_{i=1} ^{N} \{ \Phi(x_1) \cdots G_a  \Phi(x)
\cdots \Phi(x_N) \} \delta(x-x_i)
\eea
}
{ Therefore for given (small) $\eta$: }
{
\bea
&&\frac{\partial}{\partial x^\mu} < j_a ^\mu (x) \Phi(x_1) \cdots \Phi(x_N)>\cr \nonumber \\
&& = i \sum_{i=1} ^{N}  \delta(x-x_i) < \Phi(x_1) \cdots G_a \Phi(x)
\cdots \Phi(x_N) >
\eea
}
{ This is the Ward identity corresponding to current $j_a ^\mu (x)$.}
\vskip 1cm
\begin{center}
{\large $\bullet$ Ward identity corresponding to the conformal invariance}
\end{center}
{ We know that the Stress-Tensor ( $T_{\mu \nu}$) is conserved
current due to the invariance of $S$ under transformation $ x'^\mu
= x^\mu + \epsilon^\mu$ with constant $\epsilon$`s. It`s
properties are: 1) $T_{\mu \nu } = T_{\nu \mu}$ and 2) $
\partial_\mu T^{\mu \nu } = 0 $. Conserved charges are $ P^\nu =
\int d^{d-1}x T^{0 \nu}$. $P^\nu$ as an operator in Hilbert space
acting as : $ [P_\nu, \Phi] = -i \partial_\nu \Phi$. More
generally $[ Q_a, \Phi] = - i G_a \Phi$. There is another
definition of $T^{\mu \nu}$. Consider the changes in metric as $
g_{\mu \nu} \rightarrow g_{\mu \nu}' = g_{\mu \nu} + \delta g_{\mu
\nu}$, under this transformation action $S$ transforms as :} { \be
\delta S = \int d^d x \sqrt{g} T^{\mu \nu} \delta g_{\mu \nu} \ee
} { This enable us to find more restrictions of $T^{\mu \nu}$ for
conformal transformation. Suppose that theory possess Wyle
symmetry so that:
 }
{
\be
g_{\mu \nu } (x) \rightarrow \Lambda(x) g_{\mu \nu} (x)
\ee
}
{ For infinitesimal transformation, $
g_{\mu \nu} \rightarrow g_{\mu \nu} + \omega(x) g_{\mu \nu}$, or
$ \delta g_{\mu \nu}  = \omega(x) g_{\mu \nu}$. Substitute this result in eq.(39)
:}
{
\bea
\delta S &=& \int d^d x \sqrt{g} T^{\mu \nu} \omega(x) g_{\mu \nu} \cr \nonumber \\
&&\int d^d x \sqrt{g} T^{\mu} _{\mu} \omega(x) = 0 \cr \nonumber \\
&& \Rightarrow T^{\mu} _{\mu} = 0
\eea
}
{ This means that the stress-tensor is traceless.}
\vskip 0.25 cm
{ Now we can write the Ward identities (WI)  due to conformal
invariance. For invariance under translation:}
{
\be
\partial_\mu < T_\nu ^\mu  \Phi(N) > = \sum_{i=1} ^ N \delta(x - x_i)
\frac{\partial}{\partial x_i ^\nu } < \Phi(N) > \ee } {  for
Lorantz invariance ( its current is $j^{\mu \nu \alpha} = T^{\mu
\nu} x^\alpha - T^{\mu \alpha} x^\nu$ and its generators is $
L_{\mu \nu} = S_{\mu \nu} + i ( x_\mu \partial_\nu - x_\nu
\partial_\mu)$): { \bea &&\partial_\mu < ( T^{\mu \nu} x^\alpha -
T^{\mu \alpha} x^\nu ) \Phi(N) >=
\cr \nonumber \\
&&- \sum_{i=1} ^ N \delta(x - x_i) \{ x_i ^ \nu \partial_i ^
\alpha - x_i ^ \alpha \partial_i ^ \nu - i S_i ^{\nu \alpha} \} <
\Phi(N)> \eea } { Using the first identity:} { \bea
&& < ( T^{\mu \nu} - T^{\nu \mu} ) \Phi(N)> = \cr \nonumber \\
&& i \sum_{i=1} ^ N \delta(x - x_i)
S_i ^{\mu \nu} < \Phi(N) >.
\eea
}
{ WI corresponds to  the scaling invariance}
{
\bea
&&\partial_\mu < T_\nu ^ \mu x^\nu \Phi(N) > = \cr \nonumber \\
&& \sum_{i=1} ^ N \delta(x - x_i)
\{ (x_i ^ \nu \frac{\partial}{\partial x_i ^\nu} + \Delta_i) <\Phi(N)> \}
\eea
}
{ Also using the first identity:}
\\
{
\be
 < T_\mu ^\mu \Phi(N) > = \sum_{i=1} ^ N \delta(x - x_i) \Delta_i < \Phi(N) > .
\ee
}
\begin{center}
{\large $\bullet$ Conformal invariance in 2- dimensions}
\end{center}
{ Suppose ${\bf x} =(x^0, x^1)$ and $g_{\mu \nu} = Diag (1,1)$.
Using eq.(5) we have:}
{
\be
\partial_\mu \epsilon_ \nu + \partial_\nu \epsilon_ \mu = \frac{2}{d}
g_{\mu \nu} (\bf{\partial \cdot \epsilon})
\ee
}
{  For $ \nu \neq \mu$ and $ \mu = \nu = 1$ we have the following equations
}
{
\bea
&&\partial_1 \epsilon_2 = - \partial_2 \epsilon_1  \cr \nonumber \\
&& \partial_1 \epsilon_1 = \partial_2 \epsilon_2
\eea
}
{ Define $ z= x^0 + i x^1 $, $ \bar {z}= x^0 - i  x^1$,
and $d^2 s = dz d \bar{z}$ so that :}
{
\bea
g_{zz} &=& g_{\bar{z} \bar{z}} = 0 \cr \nonumber \\
g_{z \bar{z}} &=& g_{\bar{z} z} = \frac{1}{2}
\eea
}
{ now for $w = \epsilon_1 + i \epsilon_2$, the eqs. (48),
reduce to the condition $\partial_{\bar z} w(z, \bar z) =0 $.
This shows that the group of conformal transformations in two dimensions
is isomorphic to the (infinite-dimensional) group of arbitrary analytic
coordinate transformation $z \rightarrow w(z)$ and $\bar{z} \rightarrow \bar{w}(\bar{z})$.\\
The Mobius transformation is a subset of holomorphic
transformation which has the following expression: }
{
\be
z \rightarrow f(z) = \frac{a z + b}{cz+d} \hskip 1cm ad-bc = 1.
\ee
}
{ where $a, b, c $ and $d$ are complex numbers. One can show that the
 two transformations
$f_1$ and $f_2$ gives, $f_1 f_2 = f$, so that the parameters of
$f_1$ and $f_2$ be $ a_1,b_1,c_1,d_1$ and  $ a_2,b_2,c_2,d_2$,
respectively, the parameters of $f_3$ will be
 $ a_3 = a_1 a_2 + b_1 c_2 ,b_3 = a_1b_2 + b_1 a_2 ,c_3= a_1 a_2 + a_1 c_2
,d_3=c_1 b_2 + a_1 a_2$.  }
\\
\begin{center}
{\large $\bullet$ Holomorphic form of conformal ward identity}
\end{center}
\vskip .5cm
 In 2D we use the two-dimensional expression for generator of spin of i-th field
as $ S_{i \mu \nu} = s_i \epsilon_{\mu \nu}$ where $ \epsilon_{\mu \nu}
= - \epsilon_{\nu \mu} $. Therefore conformal Ward Identities reduces to the
following form:
{
\bea
&& \frac{\partial}{\partial x ^\mu} < T_\nu ^\mu \Phi(N) > = \cr \nonumber
&&- \sum_{i=1} ^ N \delta(x - x_i) \frac{\partial}{\partial x_i ^\nu} < \Phi(N)>
\cr \nonumber \\
&&\epsilon_{\mu \nu} < T^{\mu \nu} \Phi(N) > =
- \sum_{i=1} ^ N \delta(x - x_i) < \Phi(N) > \cr \nonumber \\
&&< T_{\mu} ^{\mu} (x) \Phi(N) > =
- \sum_{i=1} ^ N \delta(x - x_i)  \Delta_i < \Phi(N) >
\eea
}
{ Now using the following identity: }
{
\be
\delta^2 (x) = \frac{1}{\pi} \partial_{\bar z} \frac{1}{z} =
\frac{1}{\pi} \partial_{z} \frac{1}{\bar z}
\ee
}
{ we can write the conformal WI as:}
{
\bea
&&2 \pi \partial_{z} < T_{z \bar z} \Phi(N) >
+   2 \pi \partial_{\bar z} < T_{zz} \Phi(N) > \cr \nonumber \\
&&= -  \sum_{i=1} ^{N} \partial_{\bar z} (\frac{1}{z - w_i})
\partial_{w_i} < \Phi(N) >,  \cr \nonumber \\
&&2 \pi \partial_{z} < T_{\bar z \bar z} \Phi(N) >
+   2 \pi \partial_{\bar z} < T_{z \bar z} \Phi(N) > \cr \nonumber \\
&&= -  \sum_{i=1} ^{N} \partial_{\bar z} (\frac{1}{z - w_i})
\partial_{\bar {w_i}} < \Phi(N) >,  \cr \nonumber \\
&&2  < T_{z \bar z} \Phi(N) >
+   2  < T_{{\bar z} z} \Phi(N) > \cr \nonumber \\
&&= -  \sum_{i=1} ^{N} \delta(x - x_i) \Delta_i < \Phi(N) >,  \cr \nonumber \\
&&- 2  < T_{z \bar z} \Phi(N) >
+   2  < T_{{\bar z} z} \Phi(N) > \cr \nonumber \\
&&= -  \sum_{i=1} ^{N} \delta(x - x_i) s_i < \Phi(N) >,
\eea
}
{ Using the last equations:}
{
\bea
&&2 \pi  < T_{\bar z  z} \Phi(N) >
 \cr \nonumber \\
&&= -  \sum_{i=1} ^{N} \partial_{\bar z} (\frac{1}{z - w_i})
h_i  < \Phi(N) >,  \cr \nonumber \\
&&2 \pi  < T_{z \bar z } \Phi(N) >
 \cr \nonumber \\
&&= -  \sum_{i=1} ^{N} \partial_{ z} (\frac{1}{ \bar z - \bar w_i})
\bar {h_i}  < \Phi(N) >,
\eea
}
{  where $h= \frac{1}{2} ( \Delta + s)$ and
 $\bar h= \frac{1}{2} ( \Delta - s)$. Using the above equation
 we can rewrite the eq.(53) as:}
{
\bea
&& \partial_{\bar z} < T \Phi(N) > = \cr \nonumber \\
&&\partial_{\bar z} \{   \sum_{i=1} ^{N}  \frac{ 1}{z - w_i}
\partial_{w_i} < \Phi(N) >  + \frac{h_i}{(z-w_i)^2} < \Phi(N) > \}
\cr \nonumber \\
&& \partial_{z} < \bar{T} \Phi(N) > = \cr \nonumber \\
&&\partial_{ z} \{   \sum_{i=1} ^{N}  \frac{ 1}{\bar z - \bar w_i}
\partial_{w_i} < \Phi(N) >  + \frac{\bar h_i}
{(\bar z - \bar w_i)^2} < \Phi(N) > \}
\eea
}
\\
{ where $T = - 2 \pi T_{zz}$ and $\bar T = - 2 \pi T_{\bar z \bar
z}$. Therefore:} \vskip .5 cm { \bea
&&< T(z) \Phi(N) > =  \cr \nonumber \\
&&  \sum_{i=1} ^{N} \{ \frac{1}{z-w_i}
\partial_{w_i} < \Phi(N) > + \frac{h_i}{(z-w_i)^2} <\Phi(N)> \} + \cdots
\eea
}
{ Also same equation holds for $\bar T$ (replace $z$ with $\bar z$).
This gives us the ${\it Operator}$ ${\it  Product}$ of $T$ and $\Phi$. For example
consider $N=1$ and find that :}
{
\bea
T(z) \Phi(w) = \frac{1}{z-w} \partial_w \Phi(w) + \frac{h}
{ (z - w)^2 } \Phi + \cdots
\eea
}
In the next section we will start from  the OPE
of $T$ and $\Phi$ and introduce the LCFT.

\begin{center}
{\large $\bullet$ Correlation Functions in Two-Dimensions }
\end{center}
\vskip 0.5 cm
{ Let us now find $\delta_\epsilon \Phi(x)$. Using the eq.(36):}
{
\bea
&&\delta_\epsilon < \Phi(N) > = \int _M d^2 x \partial_\mu < T ^{\mu \nu } (x)
\epsilon_{\nu} (x) \Phi(N) >  \cr \nonumber \\
&& \frac{i}{2}  \oint _c  \{ - dz
< T^{\bar z \bar z} \epsilon_{\bar z} \Phi(N) >
+ d{\bar z}
< T^{ z  z} \epsilon_{z} \Phi(N) > \}
\eea
}
{ or}
{
\bea
\delta_{\epsilon \bar \epsilon} < \Phi(N) > = - \frac{1}{2 \pi i} \oint_c dz < T(z) \Phi(N)>
+ C.C
\eea
}
{ Using the operator product of $T(z)$ and $\Phi(w)$:}
{
\bea
\delta_\epsilon < \Phi(N) > = -  \sum_{i=1} ^{N} (\epsilon(w_i) \partial_{w_i}
+ \partial_{w_i} \epsilon(w_i) h_i ) < \Phi(N) >
\eea
}
{ Therefore for holomorphic part:}
{
\bea
\delta_\epsilon \Phi(z) =
 - \epsilon \partial_z \Phi(z) - h \Phi(z) \partial_z \epsilon
\eea
}
{  For infinitesimal transformation
$ \epsilon(z) = a + b z + c z^2 $:}
{
\bea
 &&\sum_{i=1} ^{N}  \partial_{w_i} < \Phi(N) > = 0 \cr \nonumber \\
 &&\sum_{i=1} ^{N}  ( w_i \partial_{w_i}  + h_i ) < \Phi(N) > = 0 \cr \nonumber \\
 &&\sum_{i=1} ^{N}  ( w_i^2  \partial_{w_i}  +  2 w_i h_i ) < \Phi(N) > = 0
\eea
}
{ One can solve the above equation and find that:}
{
\bea
< \Phi_1 (z_1, \bar z_1 ) \Phi_2 (z_2, \bar z_2) > = \delta_{h_1,h_2} && \frac{c_{1,2}}{ (z_1 - z_2) ^{ 2 h}
(\bar z_1 - \bar z_2) ^ {2 \bar h} }
\eea
}

\bea
&& < \Phi_1 (x_1) \Phi_2 (x_2) \Phi_3 (x_3) > = \cr \nonumber \\
&&\frac{c_{1,2,3}}{
x_{1,2} ^{h_1 + h_2 - h_3}
x_{2,3} ^{h_2 + h_3 - h_1}
x_{1,3} ^{h_1 + h_3 - h_2}}
\eea

{
\bea
&&< \Phi_1 (x_1) \Phi_2 (x_2) \Phi_3 (x_3) \Phi_4 (x_4) > = \cr \nonumber \\
&& f( \frac{ x_{1,2}x_{3,4}}{x_{1,3} x_{2,4}},
 \frac{ x_{1,2}x_{3,4}}{x_{2,3} x_{1,4}}, \cdots) \Pi_{i<j}
 ^4 x_{i,j}^ { h/3 - h_i - h_j}
\eea
}
{ where $ h = \sum_{i=1} ^4 h_i$. Note that in four point function
 unknown function
$f$ has only one crossing ratio, because, if one define
$ \eta = \frac{ z_{1,2}z_{3,4}}{z_{2,3} z_{1,4}}$, therefore will find that,
$ \frac{ z_{1,2}z_{3,4}}{z_{1,4} z_{2,3}} = \frac{\eta} {1 - \eta  } $
and $ \frac{ z_{1,4}z_{2,3}}{z_{1,3} z_{2,4}} = 1 - \eta $.
So $f$ is function of $ \eta $ and $\bar \eta$.\\

{ In the end of this sub-section let us introduce the central charge $c$
via the OPE of $T$ and $T$. For instance consider the
following actions in two dimensions;}
{
\bea
&&S_1 = \frac{1}{2 } g \int d^2x \partial_\mu \phi \partial^\mu \phi \cr \nonumber \\
&&S_2 = \frac{1}{2 } g \int d^2x \psi^{+} \gamma^0 \gamma^\mu
\partial_\mu \psi \eea } {it can be shown that  :} { \bea T(z)
T(w) = \frac{ c/2} { (z-w)^4} + \frac{2 T(w)}{(z-w)^2} +
\frac{\partial T(w)}{z-w} + \cdots \eea } {  where $c =1$ and
$c=1/2$ for $S_1$ and $S_2$, respectively. This is just the
definition of central charge $c$.}

\begin{center}
{\large  $\bullet$ Transformation of Stress-Tensor}
\end{center}
{ According to definition  of stress-tensor:}
{
\bea
\delta_\epsilon \Phi = - \frac{1}{2 \pi i} \oint \epsilon(z) dz T(z) \Phi
\eea
}
{ and  for $T(w)$: }
{
\bea
&&\delta_\epsilon T(w) = - \frac{1}{2 \pi i} \oint \epsilon(z) dz T(z) T(w)
\cr \nonumber \\
&&=  - \frac{1} {12} c \partial_w ^3 \epsilon(w)
- 2 T(w) \partial_w \epsilon(w) - \epsilon(w) \partial_w T(w)
\eea
}
{ This gives the transformation of $T$ under infinitesimal
transformation $z \rightarrow z+ \epsilon(z)$.
For finite transformation ( $ z \rightarrow W(z) $):}
\\
{
\bea
T'(W) = (\frac{dW}{dz})^{-2} [ T(z) - \frac{c}{12} \{ W ; z\} ]
\eea
}
\\
{ where $  \{ W ; z \} = \frac{ d^3 W / dz^3}{dW/dz}
- 3/2 ( \frac{d^2 W / dz^2}{dW/dz})^2$, which is called the Schwartzian
derivative. \\
As an example one can check that  for Mobius transformation $W = \frac{a z+ b}{cz+d}$ (with
$ad-bc =1$) we have $ \{W;z \} =0$.\\
Also the map $ W(z) = L/{2 \pi} \ln z$ maps the plane to the strip
with periodic boundary conditions. One can show that 1) $ <
T_{plane} > = 0 $ and 2) $ < T_{cyl} > = \frac{ - c \pi^2 } {6
L^2} $. }
\begin{center}
{\large  $\bullet$ The Virasoro Algebra \& \\
Its Representation}
\end{center}
{  Expanding the stress-tensor in terms of a Laurent series as:}
{
\bea
T(z) = \sum_n z^{-n-2} L_n
\eea
}
{  with
{
\bea
L_n = \frac{1}{2 \pi i} \oint dz z^{n+1} T(z)
\eea
}
{ Using the above expansion and operator product expansion (OPE)
of $T T$:}

\bea
&&[L_n,L_m]  = \oint \frac{dw}{2 \pi i} \oint \frac{dz}{2 \pi i}
z^{n+1}{w^{m+1}} \cr \nonumber \\
&& ( \frac{c/2}{(z-w)^4} +
\frac{2 T(w)}{(z-w)^2} +
\frac{\partial T(w)}{z-w}) \cr \nonumber \\
&&=  \oint \frac{dw}{2 \pi i} ( \frac{c}{12} n (n^2-1) w ^{n+m+1} \cr \nonumber \\
&&+ 2(n+1) w^{n+m+1} T(w) + w^{n+m+2} \partial T(w)) \cr \nonumber \\
&& [L_n,L_m] = \frac{c}{12} n (n^2-1) \delta_{n+m , 0 }  + (n-m) L_{n+m}
\eea

{ These commutation relations are known as Virasoro Algebra. We
intend to create a representation of this algebra in terms of
states, so let  $| 0 >$ be a vacuum state in our field theory such
that $L_n |0 > = 0 $ for $ n =0,1,-1$. Define the state $|h> =
\Phi(0) | 0>$ for any primary field $\Phi$ of conformal weight
$h$. It can be shown that $L_0 | h> = h | h>$, so that $|h>$ are
eigenvectors of $L_0$. Also we have $ L_n | h > = 0$ for $n > 0$.
We call these states $ \bf {highest-weight}$ states in the
representation. We can obtain all states in the representation by
applying a sequence of $L_{-n}$ operators to a highest weight
state. Alternatively we can define fields:} { \bea \Phi^{-k_1 ,
\cdots ,-k_n} (z) = L_{-k_1} \cdots L_{-k_n} \Phi(z) \eea } with {
\bea L_{-k} \Phi(w) = \oint \frac{dz}{2 \pi i} \frac{ T(z)
\Phi(w)}{(z-w)^{k-1}}. \eea }

{  where  $ k_1 + \cdots + k_n = N$. The fields $\Phi^{-k_1 ,
\cdots ,-k_n} (z)$ are known as the descendent of field $\Phi (z)$
at level $N$.
  }
\\
{ For example one can check that the stress-tensor is the
descendent of identity operator $I$, i.e.  $L_{-2} I = T(0)$.}
\\
{ Now define null vector $| \xi >$ as being a linear combination of
states of the same level. Similar to primary state it is
 it must be stable under the
operation of $L_n$ and $L_0$.
{
\bea
L_n | \xi > &&= 0 , \hskip 1cm n > 0 \cr \nonumber \\
L_0 | \xi > &&= (h + N) |\xi> \eea } { For instance we can show
that the state $(L_{-2} + a L_{-1} ^2 ) \Phi$ with $a= -
\frac{3}{2(2 h +1)}$ is a null state in the level $N=2$. The
existence of null vectors gives us additional differential
equation for determining of unknown function $f(\eta)$ in
four-point correlation functions which contains the field $\Phi$.
}
\newpage
\begin{center}
\section{ \Large -Logarithmic Conformal Field Theory}
\end{center}

In an ordinary conformal field theory primary fields are the highest weights
of the representations of the Virasoro algebra. The operator product
expansion that defines a primary field  $\P (w,\bar w)$ is
\be \label{*}
T(z) \P_i (w,\bar w)={\D_i \over (z-w)^2}\P_i (w, \bar w)+{1\over (z-w)}
\d _w \P_i (w,\bar w)
\ee
\be
T(\bar z) \P_i (w,\bar w)={\bar \D_i \over (\bar z-\bar w)}\P_i
(w, \bar w)+{1\over (\bar z-\bar w)}
\d _{\bar w} \P_i (w,\bar w)
\ee
\\

where $T(z):= T_{zz}(z)$ and $\bar T(\bar z):= T_{\bar z \bar z}(\bar z)$.

The primary fields are those which transform under $z\to f(z)$ and
\\

$\bar z\to \bar f(\bar z)$ as:
\bea \label{1}
&&\P_i(z,\bar z)\to {\P}_i(z,\bar z) '= \cr \nonumber \\
&&({\d f^{-1}\over \d z})^{\D_i}
({\d {\bar f}^{-1}\over \d \bar z})^{\bar \D_i}
\P_i(f^{-1}(z),\bar f^{-1}(\bar z))
\eea
\\
It can be shown that the OPE of $T$ and $\Phi$ is equivalent to
the commutation relation between $L_n$`s and $\Phi$. Let us
introduce the radial quantization which is defined as: \bea R(
\Phi_1(z) \Phi_2 (w)) =\cases{ \Phi(z) \Phi(w) & $|w| < |z|$ \cr
\Phi(w) \Phi(z)  ,&$  |w| > |z| $\cr} \eea Now suppose we have two
function $a(z)$ and $b(z)$ (are holomorphic) and consider the
integral $\oint_w dz a(z) b(z)$. To have a well defined expression
for this integral we should consider the radial ordered of fields.
We consider two contours $c_1$ and $c_2$ so that the $c_1$ is the
circle path ( anti-clockwise) with center in the origin and its
radius is $r= |w| + \epsilon$ and $c_2$ is similar path as $c_1$ (
but clockwise) with $r=|w| - \epsilon$. Therefore, \be \oint_c dz
a(z) b(z) = \oint_{c_1} dz a(z) b(w) - \oint_{c_2} dz b(w) a(z)
\ee Define $A = \oint dz a(z)$, therefore, \be \oint_c dz a(z)
b(z) =  [ A , b(w) ] \ee From eq. (72) one finds, \be [ L_n ,
\Phi(w) ] = \frac{1}{2 \pi i} \oint dz z ^{n+1} T(z) \Phi(w) \ee
Using the OPE of $T$ and $\Phi$ \be \label{2}
[L_n,\P_i(z)]=z^{n+1}\d_z\P_i+(n+1)z^n\D_i \P_i \ee
\\

One can regard $\D_i$'s as the diagonal elements of a diagonal matrix $D$,
\be
[L_n,\P_i(z)]=z^{n+1}\d_z\P_i+(n+1)z^nD_i^j \P_j
\ee
One can however, extend the above relation for any matrix $D$, which is not
necessarily diagonal. This new representation of $L_n$  also satisfies the
Virasoro  algebra for any arbitrary matrix $D$.
 Because we have not
altered the first term in the right hand side of the equation (85),
this is still a conformal transformation.

By a suitable change of basis, one can make  $D$  diagonal or Jordanian.
If it becomes diagonal, the field theory is nothing
but the ordinary conformal field theory. The general case is that there
are some Jordanian blocks in the matrix $D$. The latter is the
case of a LCFT [7].

 Here, there arise some other fields which do not transform
like ordinary primary fields, and are called logarithmic operators.
For the simplest case, consider a two-dimensional Jordan cell.

The fields $\P$ and $\Si$ satisfy
\be \label{10}
[L_n,\P(z)]=z^{n+1}\d_z\P+(n+1)z^n\D \P
\ee
and
\be \label{11}
[L_n,\Si(z)]=z^{n+1}\d_z\Si+(n+1)z^n\D \Si+(n+1)z^n \P,
\ee
and they transform as below
\be \label{12}
\P(z)\to ({\d f^{-1}\over \d z})^{\D}\P(f^{-1}(z))
\ee
\be \label{13}
\Si(z)\to ({\d f^{-1}\over \d z})^{\D}[ \Si(f^{-1}(z))+
\log ({\d f^{-1}(z)\over \d z}) \P(f^{-1}(z)]
\ee

Note that we have considered only the chiral fields. The
logarithmic fields, however cannot be factorized to the left- and
right-handed fields. For simplicity we derive  the results for
chiral fields. The corresponding results for full fields are
simply obtained by changing \be z^{\D}\to z^{\D}{\bar z}^{\bar \D}
\ee and \be \log z\to \log \vert z\vert^2 \ee Now compare at the
relations (86,87) and (88,89); one can assume the field $\Si$ as
the derivation of the field $\P$ with respect to its conformal
weight, $\D$. Now let us consider the action of M\"obius
generators $(L_0,L_{\pm})$ on the correlation functions. Whenever
the field $\Si$ is absent, the {\it form} of the correlators is
the same as ordinary conformal field theory. By the term {\it
form} we mean that some of the constants which cannot be
determined in the ordinary conformal field theory may be fixed in
the latter case. Now we want to compute correlators containing the
field $\Si$. At first we should compute the two-point functions.
The two-point  functions of the field $\P$ is as below \be <\P(z)
\P(w)>={c\over (z-w)^{2\D}} \ee In the ordinary conformal field
theory the constant $c$ cannot be determined only with assuming
conformal invariance; to obtain it, one should know for example
the stress-energy tensor, although for $c\neq 0$ one can set it
equal to one by renormalizing the field.

Assuming the conformal invariance of the two-point function
$<\Si(z) \P(w)>$, means that acting the set $\{ L_0,L_{\pm 1}\}$
on the correlator yields zero. Action of $L_{-1}$ ensures that the
correlator depends only on the $z-w$. the relations for $L_{+1}$
and $L_0$ are as below \bea
&&[z^2\d_z+w^2\d_w +2\D(z+w)]<\Si(z)\P(w)>+ \cr \nonumber \\
&& 2z <\P(z)\P(w)>=0
\eea
\bea
&&[z\d_z+w\d_w +2\D]<\Si(z)\P(w)>+ \cr \nonumber \\
&& <\P(z)\P(w)>=0
\eea
Consistency of these two equations for any $z$ and $w$, fixes $c$ to be zero.
Then, solving the above equation for $<\Si(z) \P(w)>$ leads to
\bea
&&<\P(z) \P(w)>=0, \cr \nonumber \\
&& <\P(z) \Si(w)>={a\over (z-w)^{2\D}}
\eea
Now assuming the conformal invariance of the two-point function
$<\Si(z) \Si(w)>$, gives us a set of partial differential equation. Solving
them, we obtain
\be
<\Si(z) \Si(w)>={1\over (z-w)^{2\D}}[b-2a \log (z-w)]
\ee
Now we extend the above results to the case where Jordanian block is
$n+1$-dimensional. So there is $n+1$ fields with the same weight $\D$.
\be
[L_n,\P_i(z)]=z^{n+1}\d_z\P_i+(n+1)z^n\D \P_i+(n+1)z^n \P_{i-1},
\ee
where $\P_{-1}=0$.

All we use is the conformal invariance of the theory. From the above fields,
only  $\P_0$ is  primary.
Acting $L_{-1}$ on any two-point function of these fields, shows that
\be
<\P_i(z) \P_j(w)>=f_{ij}(z-w).
\ee
Acting $L_0$ and $L_{+1}$, leads to
\bea
&&<[L_0,\P_i(z) \P_j(0)]>= (z\d_z +2\D)<\P_i(z) \P_j(0)> \cr \nonumber \\
&&+<\P_{i-1}(z) \P_j(0)>
+<\P_i(z) \P_{j-1}(0)>=0
\eea
\bea
&&<[L_{+1},\P_i(z) \P_j(0)]>= (z^2\d_z +2z\D)<\P_i(z) \P_j(0)>\cr \nonumber \\
&&+2z<\P_{i-1}(z) \P_j(0)> =0. \eea Then it is easy to see that
\be \label{14} <\P_{i-1}(z) \P_j(0)>=<\P_i(z) \P_{j-1}(0)>. \ee
Using $\P_{-1}=0$ and the above equation, gives us the following
two-point functions. \be <\P_i(z) \P_j(w)>=0 \qquad {\rm for
}\qquad i+j<n \ee Now solving the Ward identities for $<\P_0(z)
\P_n(w)> $ among with the relation (101), leads to \be <\P_i(z)
\P_{n-i}(w)>=<\P_0(z) \P_n(w)>=a_0 (z-w)^{-2\D}. \ee The form of
the correlation function $<\P_1(z) \P_n(w)>$ is as below \be
<\P_1(z) \P_n(w)>=(z-w)^{-2\D}[a_1+b_1\log (z-w)], \ee but the
conformal invariance fixes $b_1$ to be equal to $-2a_0$. So \bea
&&<\P_i(z) \P_{n+1-i}(w)>=<\P_1(z) \P_n(w)>= \cr \nonumber \\
&&(z-w)^{-2\D}[a_1-2a_0\log (z-w)]
\quad {\rm for }\quad i>0
\eea

Repeating this procedure for the two-point functions of the other fields
$\P_i$ with $\P_n$, and knowing that they are in the following form
\be
<\P_i(z) \P_n(w)>=(z-w)^{-2\D}\sum_{j=0}^i a_{ij}(\log(z-w))^j,
\ee
gives
\be
\sum_{j=1}^i ja_{ij}(\log(z-w))^{j-1}+2\sum_{j=0}^{i-1} a_{i-1,j}
(\log(z-w))^j =0
\ee
or
$$(j+1)a_{i,j+1}+2a_{i-1,j}=0$$
So
\bea
&&a_{i,j+1}={-2\over j+1}a_{i-1,j}=\cdots = \cr \nonumber \\
&&{(-2)^{j+1}\over (j+1)!}a_{i-j-1,0}=:{(-2)^{j+1}\over (j+1)!}a_{i-j-1}
\eea
or
\be
<\P_i(z) \P_n(w)>=(z-w)^{-2\D}\sum_{j=0}^i {(-2)^j\over j!}a_{i-j}
(\log(z-w))^j,
\ee
and also we have
\bea
&&<\P_i(z) \P_k(w)>= \cr \nonumber \\
&&<\P_{i+k-n}(z) \P_n(w)>\qquad {\rm for }\qquad
i+k\geq n.
\eea
So for the case of $n$ logarithmic field, we found all the two point
functions.
The interesting points are
\\
\noindent i)some of the two-point functions become zero.
\\
\noindent ii)some of the two-point functions are logarithmic, and
the highest power of the logarithm, which occurs in the $<\P_n
\P_n>$, is $n$.

The most general case is the case where there is more than one Jordanian
block in the matrix $D$ ,or in other words, there is more than one set of
logarithmic operators. The dimension of these blocks may be equal or not
equal. Using the same procedure, one can find that
\\
\bea
&&<\P_i^I(z)\P_j^J(w)> \cr \nonumber \\
&&=\cases{ (z-w)^{-2\D}\sum_{k=0}^{i+j-n}{(-2)^k\over k!}
a_{n-k}[\log (z-w)]^k,&$i+j\geq n$\cr 0,&$i+j<n$\cr}
\eea
where   $I$ and $J$ label the Jordan cells, $n={\rm max }\{ n_I,n_J\}$
and $n_I$ and $n_J$ are the dimensions of the corresponding Jordan cells.
\\

Also note that the conformal dimensions of the cells $I$ and $J$ must be
equal, otherwise the two-point functions are trivially zero.

Now we want to consider the three-point functions of logarithmic fields.
The simplest case is the case where, besides $\P$, only one extra
logarithmic field $\Si$ exists in the theory. The three-point functions of
the fields $\P$ are the same as ordinary conformal field theory.
\bea
&&A(z_1,z_2,z_3):=<\P(z_1) \P(z_2)\P(z_3)> \cr \nonumber \\
&&={a\over (\xi_1 \xi_2\xi_3)^{\D}}
=:a f(\xi_1,\xi_2,\xi_3),
\eea
where $$\xi_i={1\over 2}\sum_{j,k}\epsilon_{ijk}(z_j-z_k).$$

If one acts the set $\{L_0,L_{\pm 1}\} $ on the three-point function
$<\Si(z_1) \P(z_2)\P(z_3)>:=B(z_1,z_2,z_3)$, the result is an inhomogeneous
partial differential equation for $B(z_1,z_2,z_3)$ where the inhomogeneous
part is $A(z_1,z_2,z_3)$. So the form of $B(z_1,z_2,z_3)$ should be as below,
\be
B(z_1,z_2,z_3)=[b+\sum b_i\log \xi_i]f(\xi_1,\xi_2,\xi_3).
\ee
Solving the above mentioned differential equations, we find the parameters
$b_i$ to be
\be
b_1=-b_2=-b_3=a.
\ee
The final result is
\be \label{**}
<\Si(z_1) \P(z_2)\P(z_3)>= [b+a\log {\xi_1\over \xi_2\xi_3}]
f(\xi_1,\xi_2,\xi_3)
\ee
If there are two fields $\Si$ in the three-point function, one can write it
in the following form
\bea
&&<\Si(z_1) \Si(z_2)\P(z_3)>= \cr \nonumber \\
&&[c+\sum c_i\log \xi_i+\sum_{ij}c_{ij} \log\xi_i
\log\xi_j]f(\xi_1,\xi_2,\xi_3).
\eea
Again the Ward identities can be used to determine the parameters
$c_i$ and $c_{ij}$,
\bea
&&<\Si(z_1) \Si(z_2)\P(z_3)>=\cr \nonumber \\
&&[c-2b\log \xi_3+a
[(-{\log\xi_1\over \log \xi_2})^2+(\log\xi_3)^2]f(\xi_1,\xi_2,\xi_3).
\eea
Finally, for the correlator of three $\Si$'s we use
\bea
&&<\Si(z_1) \Si(z_2)\Si(z_3)>=
[d+d_1D_1+d_2D_2+ \cr \nonumber \\
&& d'_2D_1^2+d_3D_3+d'_3D_1D_2
+d'_3D_1^3] f(\xi_1,\xi_2,\xi_3)
\eea
where
\be
D_1:=\log (\xi_1\xi_2\xi_3)
\ee
\be
D_2:=\log\xi_1\log \xi_2+\log\xi_2\log \xi_3+\log\xi_1\log \xi_3
\ee
\be
D_3=\log \xi_1 \ \log \xi_2 \ \log \xi_3.
\ee
This is the most general symmetric up to third power logarithmic function
of the relative positions.
Using the Ward identities, this three-point function is calculated to be
\bea
&&<\Si(z_1) \Si(z_2)\Si(z_3)>=
 [d-cD_1+4bD_2-bD_1^2 \cr \nonumber \\
&&+8aD_3-4aD_1D_2+aD_1^3]
f(\xi_1,\xi_2,\xi_3)
\eea
Now there is a simple way to obtain these correlators.
Remember of the relation between $\P(z)$ and $\Si(z)$
\be
\Si(z)={\d\over \d\D}\P(z).
\ee
Consider any three-point function which contains the field $\Si$.
This correlator is related to another correlator which has a $\P$
instead od $\Si$ according to
\bea
&&<\Si(z_1)A(z_2)B(z_3)> \cr \nonumber \\
&&={\d\over \d\D}<\P(z_1)A(z_2)B(z_3)>,
\eea
To be more exact, the left hand side satisfies the Ward identities if the
right hand side does so.
\\
But the three-point function for ordinary fields are known. So it is enough
to differentiate it with respect to the weight $\D$. Obviously, a
logarithmic term appears in the result. In this way one can easily obtain
the above three-point functions. In fact instead of solving certain partial
differential equations, one can easily differentiate with respect to the
conformal weight. This method can also be used when there are $n$ logarithmic
fields. To obtain the three- point function containing the field $\P_i$,
one should write the three-point function, which contains the
field $\P_0$, and then differentiate it $i$ times with respect to $\D$.

Note that in the first three-point function, there may be more than
one field with the same conformal weight $\D$. Then one must treat the
conformal weights to be independent variables, differentiate with
respect to one of them, and finally put them equal to their appropriate
value.
Second, there are some constants, or unknown functions in the case of
more than three-point functions, in any correlator. In differentiation
with respect to a conformal weight, one must treat these {\it formally} as
functions of the conformal weight as well.

As an example consider
\bea
&&<\P(z_1) \P(z_2)\P(z_3)>= \cr \nonumber \\
&&{a\over (\xi_1)^{\D_2+\D_3-\D_1}
(\xi_2)^{\D_3+\D_1-\D_2}(\xi_3)^{\D_2+\D_1-\D_3}}.
\eea
Differentiate with respect to $\D_1$, and then put $\D_1=\D_2=\D_3$,
and ${\d a\over \d \D_1}=b$. This is (\ref{**}).
This method can be used for any $n$-point function:

\bea
&&<\P_i(z_1)\cdots A(z_{n-1})B(z_n)> = \cr \nonumber \\
&&{\d^i\over {\d\D}^i}<\P_0(z_1)
\cdots A(z_{n-1})B(z_n)>,
\eea
provided one treats the constants and functions of the correlator as
functions of the conformal weight.
\begin{center}
{ \large $\bullet$ Logarithmic Conformal Field Theory With
Continuous Weights}
\end{center}


In the previous subsection assuming conformal invariance
 we have explicitly calculated two-
and three-point functions for the case of more
than one set of logarithmic fields when their
conformal weights belong to a discrete set. Regarding logarithmic fields
{\it formally} as derivations of ordinary fields with respect to their
conformal dimension, we have calculated n-point functions containing
logarithmic fields in terms of those of ordinary fields.
Here, we want to consider logarithmic conformal field
theories with continuous
weights [72]. The simplest example of such theories is the free field theory.


In the last part it is shown that if there are quasi-primary fields in a
conformal field theory, it causes logarithmic terms in the correlators of
the theory. By quasi-primary fields, it is meant a family of operators
satisfying

\be \label{1}
[L_n,\P^{(j)}(z)]=z^{n+1}\partial_z\P^{(j)}(z)+(n+1)z^n\D \P^{(j)}(z)+
(n+1)z^n\D \P^{(j-1)}(z),
\ee
where $\D$ is the conformal weight of the family. Among the fields $\P^{(j)}$,
the field $\P^{(0)}$ is primary.
It was shown that one can interpret the fields $ \P^{(j)}$, {\it formally},
as the $j$-th derivative of a field with respect to the conformal weight:
\be
\P^{(j)}(z)={1\over j!}{\d ^j\over \d \D^j}\P^{(0)}(z),
\ee
and use this to calculate the correlators containing $\P^{(j)}$ in terms of
those containing $\P^{(0)}$ only. The transformation relation (\ref{1}),
and the symmetry of the theory under the transformations generated by
$L_{\pm 1}$ and $L_0$, were also exploited to obtain two-point functions
for the case where conformal weights belong to a discrete set.
There were two features in two point functions. First, for two families
$\P_1$ and $\P_2$, consisting of $n_1+1$ and
$n_2+1$ members, respectively, it was shown that the correlator
$<\P^{(i)}_1\P^{(j)}_2>$ is zero unless $i+j\geq {\rm max} (n_1, n_2)$.
(It is understood that the conformal weights of these two families are equal.
Otherwise, the above correlators are zero.) Another point was that
one could not use the derivation process with respect to the conformal
weights to obtain the two-point functions of these families from
$<\P_1^{(0)}\P_2^{(0)}>$, since the correlators contain a multiplicative term
$\delta_ {\D_1, \D_2}$, which can not be differentiated with respect to the
conformal weight.

Now, suppose that the set of conformal weights of the theory is a
continuous subset of the real numbers. First, reconsider the
arguments resulted to the fact that $<\P_1^{(i)}\P_2^{(j)}>$ is
equal to zero for $i+j\geq {\rm max} (n_1, n_2)$. These came from
the symmetry of the theory under the action of $L_{\pm 1}$ and
$L_0$. Symmetry under the action of $L_{-1}$ results in \be
<\P_1^{(i)}(z)\P_2^{(j)}(w)>=<\P_1^{(i)}(z-w)\P_2^{(j)}(0)>=:A^{ij}(z-w).
\ee We also have \be \label{4}
<[L_0,\P_1^{(i)}(z)\P_2^{(j)}(0)]>=(z\partial +\D_1
+\D_2)A^{ij}(z) +
                                   A^{i-1,j}(z) +A^{i,j-1}(z)=0,
\ee
and
\be
<[L_1,\P_1^{(i)}(z)\P_2^{(j)}(0)]>=(z^2\partial +2z\D_1 )A^{ij}(z) +
                                   2zA^{i-1,j}(z) =0.
\ee These show that \be \label{6} (\D_1 -\D_2)A^{ij}(z) +
A^{i-1,j}(z) -A^{i,j-1}(z)=0. \ee If $\D_1 \ne \D_2$, it is easily
seen, through a recursive calculation, that $A^{ij}$'s are all
equal to zero. This shows that the support of these correlators,
as distribution  of $\D_1$ and $\D_2$, is $\D_1 -\D_2=0$. So, one
can use the ansatz,
 \be \label{7} A^{ij}(z)=\sum _{k\geq 0}
A^{ij}_k(z)\delta^{(k)}(\D_1 -\D_2). \ee Inserting this in
(\ref{6}), and using $x\delta^{(k+1)}(x)=-(k+1) \delta^{(k)}(x)$,
it is seen that \be \label{8} \sum_{k\geq 0}[-(k+1)A^{ij}_{k+1}(z)
+ A^{i-1,j}_k(z) -A^{i,j-1}_k(z)] \delta^{(k)}(\D_1 -\D_2)=0, \ee
or \be \label{9} (k+1)A^{ij}_{k+1}(z)= A^{i-1,j}_k(z)
-A^{i,j-1}_k(z),  \qquad k\geq 0 \ee This equation is readily
solved: \be \label{10} A^{ij}_k(z)={1\over k!}\sum_{l=0}^k (^k_l)
A^{i-k+l,j-l}_0(z), \ee where $ A^{ij}_0$'s remain arbitrary. Also
note that $A^{ij}_k$'s with a negative index are zero. We now put
(\ref{7}) in (\ref{4}). This gives \be \label{11} (z\partial +\D_1
+\D_2)A^{ij}_k(z) + A^{i-1,j}_k(z) +A^{i,j-1}_k(z)=0, \ee Using
(\ref{10}), it is readily seen that it is sufficient to write
(\ref{11})  only for $k=0$ . This gives \be \label{12} (z\partial
+\D_1 +\D_2)A^{ij}_0(z) + A^{i-1,j}_0(z) +A^{i,j-1}_0(z)=0. \ee
Putting the ansatz \be A^{ij}_0(z)=z^{-(\D_1+\D_2)}\sum
^{i+j}_{m=0}\a^{ij}_m (\ln z)^m \ee in (\ref{12}), one arrives at
\be \label{14} (m+1)\a^{ij}_{m+1} + \a^{i-1,j}_m +\a^{i,j-1}_m=0,
\ee the solution to which is \be \a^{ij}_m={(-1)^m\over
m!}\sum^m_{s=0}(^m_s)\a^{i-m+s,j-s}_0. \ee From this \be
A^{ij}_0(z)=z^{-(\D_1+\D_2)}\sum ^{i+j}_{m=0} (\ln z)^m
{(-1)^m\over m!}\sum^m_{s=0}(^m_s)\a^{i-m+s,j-s}_0, \ee and \be
A^{ij}_k(z)=[{1\over k!}\sum ^k_{l=0}(-1)^l (^k_l) \sum
^{i+j-k}_{m=0} (\ln z)^m {(-1)^m\over
m!}\sum^m_{s=0}(^m_s)\a^{i-k-m+l+s,j-l-s}_0]z^{-(\D_1+\D_2)}. \ee
So we have \bea A^{ij}(z)&=&z^{-(\D_1+\D_2)}\sum _{k\geq 0}
\delta^{(k)}(\D_1 -\D_2)
[{1\over k!}\sum ^k_{l=0}(-1)^l (^k_l) \cr \nonumber \\
&&\sum ^{i+j-k}_{m=0} (\ln z)^m
{(-1)^m\over m!}\sum^m_{s=0}(^m_s)\a^{i-k-m+l+s,j-l-s}_0],
\eea
or
\be \label{19}
A^{ij}(z)=z^{-(\D_1+\D_2)}\sum _{p,q,r,s\geq 0}
{(-1)^{q+r+s}\over p!q!r!s!}\a^{i-p-r,j-q-s} (\ln z)^{r+s}
\delta^{(p+q)}(\D_1 -\D_2),
\ee
where
\be
\a^{ij}:=\a^{ij}_0.
\ee
These constants, defined for nonnegative values of $i$ and $j$, are arbitrary
and not determined from the conformal invariance only.

Now differentiate (\ref{19}) formally with respect to $\D_1$. In this process,
$\a^{ij}$'s are also assumed to be functions of $\D_1$ and $\D_2$.
This leads to
\bea
{\partial A^{ij}(z)\over \partial \D_1}&=&z^{-(\D_1+\D_2)}\sum _{p,q,r,s}
{(-1)^{q+r+s}\over p!q!r!s!}\{ {\partial \a^{i-p-r,j-q-s}\over \partial \D_1}
(\ln z)^{r+s}\delta^{(p+q)}(\D_1 -\D_2)
\cr \nonumber \\ &+& \a^{i-p-r,j-q-s} [(\ln z)^{r+s}\delta^{(p+q+1)}(\D_1 -\D_2)-
(\ln z)^{r+s+1}\delta^{(p+q)}(\D_1 -\D_2)]\},
\eea
or
\bea
{\partial A^{ij}(z)\over \partial \D_1}&=&z^{-(\D_1+\D_2)}\sum _{p,q,r,s}
{(-1)^{q+r+s}\over p!q!r!s!}(\ln z)^{r+s}\delta^{(p+q)}(\D_1 -\D_2)
\cr \nonumber \\
&&[(p+r) \a^{i-p-r,j-q-s}+{\partial \a^{i-p-r,j-q-s} \over \partial \D_1}].
\eea
Comparing this with $A^{i+1,j}$, it is easily seen that
\be
A^{i+1,j}={1\over i+1}{\partial A^{ij}\over \partial \D_1},
\ee
provided
\be  \label{24}
{\partial \a^{i-p-r,j-q-s} \over \partial \D_1}=(i+1-p-r)\a^{i+1-p-r,j-q-s}  .
\ee
Note, however, that the left hand side of (\ref{24}) is just a {\it formal
differentiation}. That is, the functional dependence of $\a^{ij}$'s on
$\D_1$ and $\D_2$ is not known, and their derivative is just another
constant. Repeating this procedure for $\D_2$, we finally arrive at
\be
\a^{ij}={1\over i!j!}{\partial^i \over \partial \D_1^i}
{\partial^j \over \partial \D_2^j}\a^{00},
\ee
and
\be
A^{ij}={1\over i!j!}{\partial^i \over \partial \D_1^i}
{\partial^j \over \partial \D_2^j}A^{00}.
\ee
These relations mean that one can start from $A^{00}$, which is simply
\be
 A^{00}(z)=z^{-(\D_1+\D_2)}\delta (\D_1 -\D_2)
 \a^{00},
\ee
and differentiate it with respect to $\D_1 $ and $\D_2$, to obtain $A^{ij}$.
In each differentiation, some new constants appear, which are denoted by
$\a^{ij}$'s but with higher indices. Note also that the definition is
self-consistent. So that this formal differentiation process is well-defined.

One can use this two-point functions to calculate the one-point functions of
the theory. We simply put $\P^{(0)}_2=1$. So, $\D_2=0$,
\be
<\P^{(0)}(z)>=\beta^0\delta (\D),
\ee
and
\be
<\P^{(i)}(z)>=\sum^i_{k=0}{\beta^{n-k}\over k!}\delta^k(\D),
\ee
where
\be
\beta^i:={1\over i!}{\d^i \beta^0\over \d \D^i}.
\ee
The more than two-point function are calculated exactly the same as in
previous subsection.

\begin{center}
{\large { $\bullet$ The Coulomb--gas model as an example of LCFT's}}
\end{center}
\noindent As an explicit example of the general formulation of the previous
section,
consider the Coulomb-gas model characterized by the action [8-9]
\be \label{31}
S={1\over 4\pi} \int \d^2 x \sqrt g [-g^{\mu \nu }(\partial_{\mu}  \P )
(\partial_{\nu} \P ) +i QR\P ],
\ee
where $\P$ is a real scalar field,  $Q$ is the charge of the theory,
$R$ is the scalar curvature of the surface and the surface itself is of
a spherical topology, and is everywhere flat except at a single point.

Defining the stress tensor as \be T^{\mu \nu}:=-{4\pi \over \sqrt
g}{\delta S\over \delta g_{\mu \nu}}, \ee it is readily
 seen that
\be T^{\mu \nu}=-(\partial^{\mu}  \P )(\partial^{\nu} \P )
+{1\over 2} g^{\mu \nu} g^{\a \beta}(\partial_{\a}  \P
)(\partial_{\beta} \P )-iQ [\p^{;\mu \nu }-g^{\mu \nu}\nabla^2\P
], \ee and \be \label{34} T(z):=T_{zz}(z)=-(\partial \p
)^2-iQ\partial^2 \p , \ee where in the last relation the equation
of motion has been used to write \be \P (z, {\bar z})=\p (z)+{\bar
\p }({\bar z}). \ee It is well known that this theory is
conformal, with the central charge \be c=1-6Q^2 \ee There are,
however, some features which need more care in our later
calculations. First, this theory can not be normalized so that the
expectation value of the unit operator become unity. In fact,
using $e^S$ as the integration measure, it is seen that \be
<1>\propto \delta (Q) \ee one can, at most, normalize this so that
\be \label{38} <1>= \delta (Q) \ee Second, $\p$ has a
$z$-independent part, which we denote it by $\p_0$. The
expectation value of $\p_0$ is not zero. In fact, from the action
(\ref{31}), \be <\p >=<\p_0 >={1\over N(Q)}\int \d \p_0 \p_0 \exp
(2iQ\p_0), \ee where $N$ is determined from (\ref{38}) and \be
<1>={1\over N}\int \d \p_0 \exp (2iQ\p_0). \ee This shows that
$N(0) =\pi $, and \be <\p_0 >={1\over 2i}[\delta ' (Q)
+{N'(0)\over N(0)}\delta (Q)] \ee More generally \be <f(\p_0
)>={1\over N}f({1\over 2i}{\d \over \d Q} ) (N<1>)
          ={1\over N}f({1\over 2i}{\d \over \d Q} ) [N\delta (Q)].
\ee

Third, the normal ordering procedure is defined as following.
One can write
\be
\p (z)= \p_0+\p_+ (z) + \p_- (z),
\ee
where $<0|\p_-(z)=0, \ \ \p_+(z)|0>=0$, and
\be
[\p_0,\p_{\pm}]=0.
\ee
The normal ordering is so that one puts all `-' parts at the left of all
`+' parts. It is then seen that
\be \label{46}
<:f[\p ]:>=<f(\p_0 )>
\ee
Here, the dependence of $f$ on $\p $ in the left hand side may be quite
complicated; even $f$ can depend on the values of $\p $ at different points.
In the right hand side, however, one simply changes $\p (z) \to \p_0$.

Now consider the two point function. From the equation of motion,
we have \be \label{47} <\p (z)\p (w)>=-{1\over 2}\ln (z-w) <1> +b
\ee we also have \be <:\p (z)\p (w):>=<\p_0^2>= -{1\over 4N} {\d^2
\over \d Q^2 }[ N\delta (Q)] \ee Note that there is an arbitrary
term in (\ref{47}), due to the ultraviolet divergence of the
theory. One can use this arbitrariness, combined with the
arbitrariness in $N(Q)$, to redefine the theory as \be \label{48}
\p (z)\p (w)=:-{1\over 2}\ln (z-w) +:\p (z)\p (w):, \ee and \be
<f(\p_0)>:=f({1\over 2i} {\d \over \d Q})\delta (Q) \ee these
relations, combined with (\ref{46}) are sufficient to obtain all
of the correlators. One can, in addition, use (\ref{34}) (in
normal ordered form) to arrive at \be \label{51} T(z) \p
(w)={\partial _w \p\over z-w}-{iQ/2\over (z-w)^2}+{\rm r.t.}, \ee
and \be T(z) T(w)={\partial _w T\over z-w}-{2T(w)\over (z-w)^2}+
{(1-6Q^2)/2\over (z-w)^4}. \ee Eq. (\ref{51}) can be written in
the form \be [L_n,\p (z)]=z^{n+1}\partial \p -{iQ\over 2}(n+1)z^n.
\ee {\it This shows that the operators $\p $ and $1$ are a pair of
logarithmic operators with $\D =0$ (in the sense of first part of
section 3)}. Note that $ <1>$ is equal to $\delta(Q)$.
 One can easily show that
\be \label{54} T(z) :{\rm e}^{i\a \p (w)}: \ ={\partial _w :{\rm
e}^{i\a \p (w)}:\over z-w}- {\a (\a +2Q)/4\over (z-w)^2}:{\rm
e}^{i\a \p (w)}: + {\rm r.t.}, \ee which shows that $:{\rm e}^{i\a
\p }:$ is a primary field with \be \D_{\a}={\a (\a +2Q) \over 4}
\ee To this field, however, there corresponds a quasi conformal
family (pre--logarithmic operators [55]), whose members are
obtained by explicit differentiation with respect to $\a$ ($\a$ is
not the conformal weight but is a function of it): \be W_{\a
}^{(n)}:\ =\ :\p^n {\rm e}^{i\a \p }:\ = (-i)^n {\d \over \d \a^n
} :{\rm e}^{i\a \p }:. \ee

To calculate the correlators of $W$'s, it is sufficient to calculate
$<W_{\a_1}^{(0)}\cdots W_{\a_k}^{(0)}>$.

One has, using Wick's theorem and (\ref{48}),
\be
\Pi^k_{j=1} :{\rm e}^{i\a_j \p (z_j)}: \ ={\rm e}^{1/2 \sum_{1\leq i<j\leq k}
\a_i \a_j\ln (z_i-z_j)}:{\rm e}^{i\sum_{j=1}^k\a_j \p (z_j)}:
\ee
From  this using (\ref{46}) and (\ref{48}), we have
\bea
<\Pi_{j=1}^kW_{\a_j}^{(0)}(z_j)>&=&
[\Pi_{ \scriptstyle{1\leq i<j\leq k}}(z_i-z_j)^{\a_i \a_j\over 2}]
{\rm e}^{1/2\sum_{j=1}^k\a_j{\d \over \d Q} }\delta (Q) \cr \nonumber \\
&=& [\Pi_{\scriptstyle{1\leq i<j\leq k}}(z_i-z_j)^{\a_i \a_j\over 2}]
\delta (Q+{1\over 2} \sum_{j=1}^k \a_j).
\eea

Obviously, differentiating with respect to any $\a_i$, leads to logarithmic
terms for the correlators consisting of logarithmic fields $W_{\a}^{(n)}$.
The power of logarithmic terms is equal to the sum of superscripts of the
fields $W_{\a}^{(n)}$.

\begin{center}
{ \large $\bullet$ Logarithmic Conformal Field Theory in $d$-dimensions }
\end{center}

In the previous subsection we discussed the LCFT in 2-dimensions.
Generalization for arbitrary dimension $d$ has been given in [68].
We have dealt with two dimensional conformal field theory relying
heavily on the underlying Virasoro algebra, and have described how
the appearance of logarithmic singularities is related to the
modification of the representation of the Virasoro algebra. In
this subsection  we try to understand LCFT's in the context of
d-dimensional conformal invariance.\newline As is well known, one
of the basic assumptions of conformal field theory is the
existence of a family of operators, called scaling fields, which
transform under scaling $S: x\rightarrow x'=\ll x$, simply as
follows: \be  \l{scalet} \phi (x) \rightarrow \phi '(x')=\ll
^{-\Delta} \phi (x), \ee where $\Delta$ is the scaling weight of
$\phi (x)$. It is also assumed that under the conformal group,
such fields transform as, \be \l{conft} \phi (x) \rightarrow \phi
'(x')=\parallel {{\p x'}\o {\p x}}\parallel ^{ {{-\Delta} \o {d}}}
\phi (x), \ee where $d$ is the dimension of space and $\parallel
{{\p x'}\o{\p x}}\parallel $ is the Jacobian of the
transformation. Equation (\ref{conft}) which encompasses eq.
(\ref{scalet}) defines the transformation of the quasi-primary
fields. For future use we note that the Jacobian equals $\ll ^d$
for scaling transformation and $\parallel x\parallel ^{-2d}$ for
the Inversion transformation $I: x\rightarrow
 x'={{x}\o
{{\parallel x\parallel}^2}}$, being unity for the other elements of the
conformal group.
Combination of (\ref{conft}) with the definition of symmetry of the correlation
functions, i.e.:
\be  \l{invarcor}
<\phi '_1(x'_1)\cdots\phi '_N(x'_N)>=
<\phi _1(x'_1)\cdots\phi _N(x'_N)>,
\ee
allows one to determine the two and the three point functions up to a
constant and the four point function up to a function of the cross ratio.
\newline
{\it It's precisely the assumption that scaling fields constitute irreducible
representations
of the scaling transformation, which imposes power law singularity on the
correlation functions}. As we will see, relaxing this assumption, one naturally
 arrives
at logarithmic singularities. It also leads to many other
peculiarities, in the relation between correlation functions. To
begin with, we consider a multiplet of fields, \be \Phi = \left
(\begin{array}{c} \phi _1 \\ \phi _2 \\ \vdots \\ \phi _n
\end{array} \right ),
\ee
 and note that  under scaling $x\rightarrow \ll x$, the
most general form of the transformation of $\Phi$ is,
\be \l{mt}
\Phi (x)\rightarrow \Phi '(x')=\ll ^{T'}\Phi (x)
\ee
where $T'$ is an arbitrary matrix. More generally, we replace (\ref{mt}) by,
\be \l{confgt}
\Phi (x) \rightarrow \Phi '(x')=\parallel {{\p x'}\o {\p x}}\parallel ^T
\Phi (x).
\ee
where $T$ is an $n\times n$ arbitrary matrix. When $T$ is diagonalizable, one
arrives at ordinary scaling fields by redefining $\Phi$, so that all the fields
transform as $1$-dimensional representation. Otherwise, following [68] we
assume that $T$ has Jordan form,
\be  \l{t}
T=\pmatrix { {{-\Delta}\o {d}}&0&\cdots&0\cr
1&{{-\Delta}\o {d}}&\cdots&0\cr
0&1&\ddots&0\cr
0&\cdots&1&{{-\Delta}\o {d}} }.
\ee
Rewriting $T$ as ${{-\Delta}\o{d}}1+J$, where $J_{ij}=\delta _{i-1,j}$,
eq. (\ref{confgt}) can be written in the form,
\be \l{confgtlam}
\Phi (x) \rightarrow \Phi '(x')=\parallel {{\p x'}\o {\p x}}\parallel ^{{{-
\Delta} \o {d}}}\Lambda _x
\Phi (x),
\ee
where $\Lambda _x=\parallel {{\p x'}\o {\p x}}\parallel ^{J}$ is a lower
triangular matrix of the form,
\be \l{lambda}
{(\Lambda _x)}_{ij}={{{\{\ln \parallel {{\p x'}\o{\p x}}\parallel\}}^{i-j}}\o
{(i-j)!}},\qquad  {(\Lambda _x)}_{ii}=1,
\ee
i. e. for $N=2$ we have,
\bea \l{N2}
\phi _1'(x')&=&{\parallel {{\p x'}\o {\p x}}\parallel}^{{{-\Delta} \o {d}}}
\phi _1(x),\nn\\
\phi _2'(x')&=&{\parallel {{\p x'}\o {\p x}}\parallel}^{{{-\Delta}
\o {d}}} \big( \ln \parallel {{\p x'}\o {\p x}}\parallel \phi
_1(x)+\phi _2(x) \big). \eea An important point is that the top
field $\phi _1(x)$ {\it always} transform as an ordinary
quasi-primary field. A most curious property of the transformation
(\ref{confgtlam}) is that each field $\phi _{k+1}$ transforms as
if it is a formal derivative of $\phi _k$ with respect to
${{-\Delta} \o {d}}$, \be  \l{der} \phi _{k+1}(x)={1 \o k}{{\p}\o
{\p ({{-\Delta }\o {d}})}}\phi _k(x). \ee This formal relation
which determines the transformation of all the fields of a Jordan
cell from that of the top field $\phi _1$, essentially means that
with due care, one can determine the correlation functions of the
lower fields from those of the ordinary top fields simply by
formal differentiation. Also one can find the two-point,
three-point correlation functions of fields for jordan -cell of
rank $r$. For example we already know by standard arguments that
the two point function of the top fields $\phi _\alpha$ and $\phi
_\beta$ belonging to two different Jordan cells $(\Delta _\alpha
,n)$ and $(\Delta _\beta ,m)$ vanishes, i.e.: \be \l{zero} <\phi
_\alpha (x)\phi _\beta (y)>={{A {\delta _{{\Delta _\alpha},{\Delta
_\beta}}}}\o{\parallel x-y\parallel^{2\Delta _\alpha}}}. \ee Due
to the observation (\ref {der}), it follows that the two point
function of all the fields of two different Jordan cells with
respect to each other vanish. Therefore we can calculate the two
point function of the fields within the same Jordan cell. As we
will see logarithmic conformal symmetry gives many interesting and
unexpected results in this case. Let's denote the matrix of two
point functions  $<\phi _i (x)\phi _j (y)>$ for all $\phi _i,\phi
_j \in (\Delta ,n)$ by $G(x,y)$, then from rotation and
translation symmetries, this matrix should depends only on
$\parallel x-y \parallel$. From scaling symmetry and using
(\ref{conft}) and (\ref{invarcor}), we have, \be \l{ss} \Lambda
G(\parallel x-y \parallel )\Lambda ^t=\ll ^{2\D} G(\ll
\parallel x-y
\parallel ),
\ee
where $\Lambda =\ll ^{dJ}$, and from inversion symmetry, we have,
\be \l{is}
\Lambda _xG(\parallel x-y \parallel )\Lambda _y^t=\parallel x-y\parallel
^{-2\D _\alpha}
G({{\parallel x-y\parallel}\o{\parallel x\parallel \parallel y\parallel}}),
\ee
where \bea \l{lamx}\Lambda _x &=&\parallel x\parallel ^{-2dJ},\nn\\
\Lambda _y &=&\parallel y\parallel ^{-2dJ}.\eea
Defining the matrix $F$ as $G(\parallel x-y\parallel )={{F(\parallel x-y
\parallel )}\o {
{\parallel x-y \parallel }^{2\D}}}$, we will have from (\ref {ss})
and (\ref {is}), \be \l{ssf} \Lambda F(\parallel x-y \parallel
)\Lambda ^t=F(\ll \parallel x-y \parallel ), \ee and \be \l{isf}
\Lambda _xF(\parallel x-y \parallel )\Lambda _y^t= F({{\parallel
x-y\parallel}\o{\parallel x\parallel \parallel y\parallel}}). \ee
For every arbitrary $\ll$, we now choose the points $x$ and $y$
such that $\parallel x\parallel =\ll ^{{{-1}\o{4}}}$ and
$\parallel y\parallel =\ll ^{{{-3}\o{4}}}$. {\it It should be
noted that in this way by varying $\ll$, we can span all the
points of space}. From (\ref {lamx}), we will have $\Lambda
_x=\Lambda ^{{1\o 2}}$ and $\Lambda _y=\Lambda ^{{3\o 2}}$,
therefore eq. (\ref {ssf}) turns into, \be \l{isff} \Lambda ^{{1
\o 2}}F(\parallel x-y \parallel ){\big ( \Lambda ^{{3\o 2}} \big )
}^t= F(\ll \parallel x-y\parallel).\ee Combining (\ref {ssf}) and
(\ref {isff}) and using invertibility of $\Lambda$, we arrive at,
\be \l{f} F=\Lambda ^{{1 \o 2}}F{\big (\Lambda ^t\big )}^{{-1}\o
2}, \ee by iterating (\ref{f}), we will have $F=\Lambda F\Lambda
^t$, and by rearranging, we have,
 \be \l{ff} F\Lambda ^t=\Lambda
F. \ee Expanding $\Lambda $ in terms of power of $\ln \ll$ as
$\Lambda =1+(d\ln \ll )J+{{(d\ln \ll )^2}\o {2!}}J^2+\cdots $ and
comparing both sides, we arrive at \be \l{fff} F{\big (J^t\big
)}^k={\big ( J\big )}^kF,\qquad k=1,2,\cdots ,n-1. \ee Since
${\big ( J^k\big )}_{ij}=\delta _{i,j+k}$, we will have from (\ref
{fff}), \be \l{ffff} F_{i,j-k}=F_{i-k,j}, \ee which means that on
each opposite diagonal of the matrix $F$, all the correlations are
equal. Moreover from $FJ=JF$, one obtains, \be \l{f5} \sum
_{l=1}^n F_{il}\delta _{j,l+1}=\sum _{l=1}^n\delta _{i,l+1}F_{lj},
\ee which means that if $j=1$ and $1<i\leq n$, then $F_{i-1,j}=0$,
or \be \l{finalf} F_{ij}=0 \qquad {\rm for}\qquad j=1 \qquad{\rm
and}\qquad 1\leq i\leq n-1.\ee Combining this with (\ref {ffff}),
we find that all the correlations above the opposite diagonal are
zero. In order to find the final form of $F$, we use eq. (\ref
{ssf}) again, this time in infinitesimal form, let $\Lambda
=1+\alpha J+o(\alpha ^2)$ where $\alpha =d\ln \ll $, then from
$\Lambda F(x)\Lambda ^t=F(\ll x)$, we have, \be \l{infi} d\big
(JF+FJ^t\big )=x{{dF}\o{dx}}. \ee Due to the property (\ref
{ffff}) only the last column of $F$ should be found, therefore
from (\ref {infi}) we obtain, \bea \l{lc}
x{{dF_{1,n}}\o {dx}}&=&dF_{1,n-1}\equiv 0,\nn\\
x{{dF_{i,n}}\o {dx}}&=&2dF_{i,n-1},\qquad {\rm if}\quad i>1
\eea
which upon introducing the new variable $y=2d\ln x$ gives,
\be \l{fs}
F_{1,n}=c_1,\quad
F_{2,n}=c_1y+c_2,\quad
F_{3,n}={ 1\o 2}c_1y^2+c_2y+c_3,\quad
{\rm etc.}
\ee
with the recursion relations,
\be \l{rec}
{{dF_{i,n}} \o {dy}}=F_{i-1,n}.
\ee
Thus we have arrived at the final form of the matrix $F$, which is as follows:
\be \l{fin}
F=\pmatrix{0&\cdots&0&0&g_0\cr
           0&\cdots&0&g_0&g_1\cr
           0&\cdots&g_0&g_1&g_2\cr
           \vdots&\vdots&\vdots&\vdots&\vdots\cr
           g_0&\cdots&g_{n-2}&g_{n-1}&g_n
           },
\ee
where each $g_i$ is a polynomial of degree $i$ in $y$, and $g_i={{dg_{i+1}}\o
{dy}}$. All the correlations depend on the $n$ constants $c_1,\cdots ,c_n$,
which remain undetermined. We have checked (as the reader can check for
the single $n=2$ case) that inversion symmetry puts no further restrictions
on the constants $c_i$.

The observation that the transformation properties of the members
of a Jordan cell are as of the  formal derivative of the top field in the
cell, allows one to determine the correlation functions of all the fields within
a single or different Jordan cells, once the correlation function of
the top fields are determined. As an example, from ordinary CFT, we know
that conformal symmetry completely determines the three point function up to
a constant. Let $\phi _\alpha$, $\phi _\beta$ and $\phi _\gamma$ be the
top fields of three Jordan cells $(\Delta _\alpha ,l)$, $(\Delta _\beta ,m)$ and
$(\Delta _\gamma ,n)$ respectively. Therefore we know
that,
\be \l{ppp}
<\phi _\alpha (x)\phi _\beta (y)\phi _\gamma (z)>={{A_{\alpha\beta\gamma}}
\o {{\parallel x-y\parallel} ^{\Delta _\alpha +\Delta _\beta -\Delta _\gamma}
{\parallel x-z\parallel} ^{\Delta _\alpha +\Delta _\gamma -\Delta _\beta}
{\parallel y-z\parallel} ^{\Delta _\beta +\Delta _\gamma -\Delta _\alpha}}},
\ee
where the constant $A_{\alpha\beta\gamma}$ in principle depends on the weights
$\Delta _\alpha ,\Delta _\beta $ and $\Delta _\gamma$.
Denoting the second field of the cell $(\Delta _\alpha ,l)$ by $\phi _{
\alpha 1}$,\footnote {For simplicity, we have denoted the top field by $\phi$
and the second field by $\phi _1$, instead of $\phi _1$ and $\phi _2$
respectively.}
we will readily find from eq.(213) that,
\bea \l{three}
<\phi _{\alpha 1}(x)\phi _\beta (y)\phi _\gamma (z)>&=&-d {{\partial}\o{
\partial \Delta _\alpha}}<\phi _{\alpha }(x)\phi _\beta (y)\phi _\gamma (z)>
\nn\\
&=&
{
{A'_{\alpha\beta\gamma}} \o
{
{\parallel x-y\parallel} ^{\Delta _\alpha +\Delta _\beta -\Delta _\gamma}
{\parallel x-z\parallel} ^{\Delta _\alpha +\Delta _\gamma -\Delta _\beta}
{\parallel y-z\parallel} ^{\Delta _\beta +\Delta _\gamma -\Delta _\alpha}
}
}\nn\\&+&
{
{dA_{\alpha\beta\gamma}}
\o {
{\parallel x-y\parallel} ^{\Delta _\alpha +\Delta _\beta -\Delta _\gamma}
{\parallel x-z\parallel} ^{\Delta _\alpha +\Delta _\gamma -\Delta _\beta}
{\parallel y-z\parallel} ^{\Delta _\beta +\Delta _\gamma -\Delta _\alpha}
}} \cr \nonumber \\
&& \ln \big ({{\parallel y-z\parallel }\o{(\parallel x-y\parallel )(\parallel
x-z\parallel )}}\big ),
\eea
where $A'_{\alpha\beta\gamma}=-d{{\partial} \o {\partial \Delta _\alpha}}
A_{\alpha\beta\gamma}$ is a new undetermined constant. For the correlation
functions
of fields within a single cell, one should then take the limit $\beta ,\gamma
\rightarrow \alpha$ in the above formula. It's not difficult to check that this
formula satisfies all the requirements demanded by conformal symmetry.
\newpage

\section{\LARGE -Disordered Systems \&
Logarithmic Conformal Field Theory}

\vskip 1cm
\begin{center}
{\large
$\bullet$ Introduction}
\end{center}
\vskip 1cm

Consider a renormalization group transformation acting on the
space of all possible couplings of a model, $\{k\}$ [1]. The
transformation has the form  $\{k'\}=R \{k\}$ where $R$  depends,
in general, on the specific transformation chosen, and in
particular, on the length rescaling parameter $b$. Suppose there
is a fixed point at $\{k\}= \{k^{*}\}$ the renormalization group
equations, linearized about the fixed point, are \be
k'_{a}-k^{*}_{a}\sim \sum_{b}T_{ab}(k_{b}-k_{b}^{*}) \ee where
$T_{ab}=\frac{\partial k'_{a}}{\partial k_{b}}|_{k=k^*}$. Denote
the eigenvalues of the matrix $T$ by $\lambda^{i}$ and its left
eigenvectors by $\{\phi^{i}\}$ so that \be
\sum_{a}\phi^{i}_{a}T_{ab}=\lambda^{i}\phi^{i}_{b} \ee Now we
define scaling variables $u_{i}\equiv
\sum_{a}\phi_{a}^{i}(k_{a}-k_{a}^{*})$, which are linear
combinations of the deviations $k_{a}-k_{a}^{*}$ from the fixed
point which transform multiplicatively near the fixed point \bea
u_{i}&=&
\sum_{a}\phi_{a}^{i}(k'_{a}-k_{a}^{*})=\sum_{a,b}\phi_{a}^{i}T_{ab}
(k_{a}-k_{a}^{*})\nonumber\\
&=&\sum_{b}\lambda^{i}\phi_{b}^{i}(k_{b}-k_{b}^{*})=\lambda^{i}u_{i}
\eea It is convenient to define the quantities $y_{i}$ by
$\lambda^{i}=b^{y_{i}}$, the $y_{i}$'s are called RG eigenvalues.
Now we can distinguish three cases,
\\
1-If $y_{i}> 0$, $u_i$ is relevant,
\\
2-If $y_{i}<0$, $u_{i}$ is irrelevant,
\\
3-If $y_{i}=0$, $u_{i}$ is marginal.
\vskip 1cm
\begin{center}
{ \large {$\bullet$ Rule of the rescaling factor $b$}}
\end{center}
Consider an infinitesimal transformation, that $b=1+\delta l$,
with $\delta l<<1$. In this case the coupling transforms
infinitesimally \be k_{a}\rightarrow k_{a}+(\frac{dk_{a}}{d
l})\delta l+O((\delta l)^{2}) \ee and the RG equation has the
differential form \be \frac{d k_{a}}{d l}=-\beta_{a}(\{k\}) \ee
where the function $\beta_{a}$ are called the RG beta-function.
Also the matrix of derivatives at the fixed point is now
$T_{ab}=\delta_{ab}+(\frac{\partial \beta_{a}}{\partial
k_{b}})\delta l$, with eigenvalues $(1+ \delta l)^{y_{i}} \sim
1+y_{i}\delta l$. Hence the $y_{i}$'s are simply the eigenvalues
of the matrix $-\frac{\partial \beta_{a}}{\partial k_{b}}$
evaluated at the zero of the beta-functions (because at the fixed
points we have $k_{a}\rightarrow k_{a}$ and beta-function must be
zero).
\begin{center}
{ \large $\bullet$ The perturbative RG}
\end{center}

When two fixed points are sufficiently close, it is then possible
to deduce universal properties at one fixed point in terms of
those at the other. Such an analysis is the basis of the
$\epsilon$-expansion and many other similar techniques. It also
allows us to describe the properties of fixed points with exactly
marginal scaling variables.

Now consider a fixed point Hamiltonian $H^*$ which is perturbed by
a number of scaling fields, so that the partition function is [1]:
\be
Z= Tr \exp\{-H^* - \sum_i g_i \sum_r a ^{x_i} \phi_i(r) \}
\ee
Let us expand this in powers of couplings $g_i$,
\bea
Z&=& Z^* \{ 1-\sum_ig_i\int<\phi_i(r) > \frac{d^dr}{a^{d-x_i}}
\cr \nonumber \\
&+&1/2\sum_{ij}g_ig_j
\int<\phi_i(r_1) \phi_j(r_2)>\frac{d^dr_1d^dr_2}{a^{2d-x_i-x_j}} \cr \nonumber\\
&-&1/{3!}\sum_{ijk}g_ig_jg_k\int<\phi_i(r_1) \phi_j(r_2)\phi_k(r_3)>
\frac{d^dr_1d^dr_2d^dr_3}{a^{3d-x_i-x_j-x_k}} \cr \nonumber \\
&+& \cdots \}
\eea
where all correlation functions are to be evaluated with respect to
the fixed point hamiltonian $H^*$.
We now implement the RG by changing the microscopic cut-off $a\rightarrow ba$
, with $b=1+\delta l$, and asking now the couplings $g_i$ should be changed
in order to preserve the partition function $Z$.
 The length $a$ appears in
three ways in eq.(221),\\
1- Explicitly , through the divisors $a^{d-x_i}$  \\
2- Implicity , we must restrict all integrals to $|r_i-r_j|>a$. \\
3-Through the dependence on the system size $L$ in the dimensionless ratio
$L/a$. \\
For the $explicit$ dependence we have:
\be
g_i\rightarrow (1+\delta l)^{d-x_i}g_{i} \sim g_i + (d-x_i)g_i\delta l
\ee
 The $implicit$ dependence will appear in the second order term.
 After changing $ a \rightarrow a(1+ \delta l)$ we may break up the integrals,
\be
\int_{|r_1 - r_2| > a(1+\delta l)} =
\int_{|r_1 - r_2| > a}
- \int_{a(1+ \delta l) >|r_1 - r_2| > a}
\ee
The first term simply gives back the original contribution to $Z$ and the second
term may be expressed using the operator product expansion
\be
\phi_i(r_1) \phi_j(r_2) =  \sum_k C_{ijk}(r_1 - r_2)  \phi_k({(r_1+r_2)}/2)
\ee
where $C_{ijk}(r_1-r_2)= \frac{c_{ijk}}{|r_1-r_2|^{x_i+x_j-x_k}}$,
($c_{ijk}$ are known as OPE coefficients) as:
\bea
&&\frac{1}{2} \sum_{i,j} c_{ijk} a ^{x_k - x_i - x_j}
\int_{ a(1+ \delta l) >|r_1 - r_2| > a}  \cr \nonumber \\
&& < \phi_k({(r_1 +r_2)}/{2}) > \frac{ d^d r_1 d^d r_2}{ a^{2d -
x_i - x_j}} \eea The integral gives a factor $S_d a^d \delta l$,
where $S_d = {2 \pi^{d/2}}/ {\Gamma(d/2)}$ ( the area of
d-dimensional sphere). This term may then be compensated  by
making the change \be g_k \rightarrow  g_k - 1 / 2 S_d \sum_{i,j}
c_{ijk} g_i g_j \delta l \ee Finally by changing the variable $
g_k \rightarrow (2/S_d) g_k$ we find the beta-function for
coupling $g_k$ as (with $ y_k = d-x_k$): \be \frac{d g_k}{d l} =
y_k g_k -  \sum_{i,j} c_{ijk} g_i g_j + \cdots \ee.

\newpage
\begin{center}
{ \Huge $\bullet$ Quenched Random Ferromagnets}
\end{center}

We consider the random bond Ising model and suppose that
 positions of the impurities
are fixed , and tracing over the only magnetic degrees of freedom
[5,60]. Let us describe these disordered systems in the continuum
limit by the following Hamiltonian, \be H=H^{0}+\int j(r) E(r)
d^{d}r \ee where $H^{0}$ is the Hamiltonian of the renormalization
group at fixed point describing the pure Ising model. The field
$j(r)$ is a quenched random variable coupled to the local energy
density $E(r)$. For simplicity we assume that the random variable
$j(r)$ is a gaussian variable which is characterized with its two
moments $< j(r) > = 0$ and $< j(r) j(r') > = g \delta( r-r') $.
The explicit distribution of $j(r)$ is, \be P( j(r) ) = N \exp( -
\frac{1}{2g} \int j(r)^2 d^d r ) \ee where $N$ is normalization
constant.

We are interested in computing the average of the free energy, or
quenched averaged correlation functions. Since the free energy is
proportional to logarithm of the partition functions $Z[j]$, we
have to compute the average of $\log Z[j]$. The replica method is
based on the identity \be \log Z[j] = \lim _{n \rightarrow 0}
\frac{Z^n[j] - 1 }{n} = \frac{d}{dn} Z^n[j] |_{n=0} \ee The
standard procedure of averaging over disorder is to introduce
replicas , i.e. , $n$ identical copies of the same model

\be Z^{n}=Tr\exp\{-\sum_{a=1}^{n}H_{a}^{0}-\int d^{d}r j(r)
\sum_{a=1}^{n} E_{a}(r)\} . \ee Averaging over the disorder gives
rise to the following effective replica Hamiltonian: \be
H_{R}=\sum_{a=1}^{n}H_{a}^{0}-g\int\sum_{a\neq b}^{n}
E_{a}(r)E_{b}(r) d^{d}r . \ee We keep only the non-diagonal terms
, since using the operator algebra of the pure system one can
absorb the diagonal terms into $H_{a}^{0}$. The replicas are now
coupled via the disorder. The scaling dimension of coupling $g$ is
$y_{g}=d-2x^{0}_{E}$ and it is relevant at the pure fixed point if
$d > 2x_{E}^{0}$. If $y_{g}$ is small , it is possible to develop
perturbative RG in powers of these variable and we can find a
random fixed point with perturbative RG.

To find the renormalization group equation for $g$ we need the
operator  product expansion of $(\sum_{a\neq b} E_{a}E_{b})$ with
itself. Since the replicas are decoupled in $H^{0}$ , we may
evaluate this using the operator product expansion of $E_{a}$ with
itself, the first few terms of which have the form \be E_{a} \cdot
E_{b}\sim \delta_{ab}+c\delta_{ab}E_{a}+\cdot\cdot\cdot, \ee where
$c$ is a coefficient whose value is fixed and universal. The
coefficient $c$ vanishes when $d=2$. This is a consequence of the
self-duality if Ising model in two dimensions [1]. Now we have \be
(\sum_{a\neq b} E_{a}E_{b})\cdot (\sum_{c\neq d} E_{c}E_{d})=
\sum_{a\neq b,c\neq d} E_{a}E_{b} E_{c}E_{d}. \ee The OPE of
$E_{a}E_{b}$ is zero (see eq.(238)) and the OPE of $E_{c}E_{d}$'s
is also zero, so we have two ways of writing the OPE. So \bea
&& \sum_{a\neq b,c\neq d} E_{a}E_{b} E_{c}E_{d} \cr \nonumber \\
&& = 2\sum_{a\neq b,c\neq d}(\delta_{bd}+c\delta_{bd}E_{b})
E_a E_c.
\eea
The first term is
\bea
&& 2\sum_{a\neq b,c\neq d}\delta_{bd} E_a E_c \cr \nonumber \\
&&= 2 \{ (n-1) \sum_{a=1} E_a^2 + (n-2) \sum_{a\neq b} E_a E_b \},
\eea
the second term can be written as;
\bea
&&2 c \sum_{a\neq b,c\neq d} \delta_{bd} E_a E_c E_b \cr \nonumber
&&= 2c \sum_{a\neq b,c\neq d} \delta_{bd} ( \delta_{ac} + c \delta_{ac} E_a) E_b
\cr \nonumber \\
&& = 2c \sum_{a\neq b,c\neq d} \delta_{bd} \delta_{ac} E_b +
2c^2 \sum_{a\neq b,c\neq d} \delta_{bd} \delta_{ac} E_a E_b \cr \nonumber \\
&& = 2 c (n-1) \sum_{a=1} E_a + 2 c^2 \sum_{a\neq b,c\neq d}
\delta_{bd} \delta_{ac} E_a E_b
 \cr \nonumber \\
&& =
 2 c (n-1) \sum_{a=1} E_a + 2 c^2 \sum_{a\neq b}
 E_a E_b
\eea
Therefore:
\bea
&&(\sum_{a\neq b} E_{a}E_{b})\cdot (\sum_{c\neq d} E_{c}E_{d}) \cr \nonumber \\
&&=
(2(n-2)+2c^{2})\sum_{a\neq b}E_{a}E_{b}\cr \nonumber \\
&&+2(n-1) \sum_{a=1} E_a^2 + 2 c(n-1) \sum_{a=1} E_a \eea The
renormalization group equation for $g$ is thus \be \frac{d
g}{dl}=y_{g}^{0}g - (2(n-2)+2c^{2})g^{2}+O(g^{3},\cdot\cdot\cdot).
\ee Therefore if one denote $\sum_{a\neq b} E_{a}E_{b}$ with
$\Phi$ so the coefficient in OPE
$\Phi\cdot\Phi=1+b\Phi+\cdot\cdot\cdot$ , is $2(n-2)+2c^{2}$.

Now Consider a scaling operator $\phi$ with a scaling dimension $x_{\phi}$
so that $\phi\Phi=b_{\phi}\phi+\cdot\cdot\cdot$ and denote coupling of $\phi$
with $t$. Now we have

\bea
\beta_{g}&=& y_{g}^{0}g-bg^{2}+\cdot\cdot\cdot \cr \nonumber \\
\beta_{\phi}&=& y_{\phi}^{0}t - 2b_{\phi}t g+ \cdot\cdot\cdot
\eea
From the above equations,

\be
g^{*}=\frac{y_{g}}{b}
\ee
Therefore using eq.(245) we obtain:

\be
y_{\phi}=\frac{\partial \beta_{\phi}}{\partial t}|_{0}=y_{\phi}^{0}-
2b_{\phi}\frac{y_{g}^{0}}{b},
\ee

and then
\be
d-x_{\phi}=(d-x_{\phi}^{0})-\frac{2b_{\phi}y_{g}^0}{b},
\ee
so

\be
x_{\phi}=x_{\phi}^{0}+\frac{2b_{\phi}y_{g}^0}{b}.
\ee

Now consider the case where $\phi$ is
 $E_t = \sum_a E_a$ and  $\tilde{E}_a= E_{a}-(\frac{1}{n})\sum_{a}E_{a}$
with $\sum_{a}\tilde{E}_{a}=0$. The combination $E_t = \sum _{a=1}
^n  E_a$ is a singlet ( symmetric under the permutation or the
replica group) and the $\tilde E_a= E_a - \frac{1}{n} \sum _{b=1}
^n E_b $ transforms according to an $(n-1)$-dimensional
representation of $S_n$. The fields $E_t$ and $\tilde{E}_{a}$ have
proper scaling dimensions. To find the scaling dimensions of new
fields we should find the OPE coefficients $ E_t \Phi$ and $
\tilde{E}_{a} \Phi$, \bea
E_t \Phi&=&(\sum_{a=1}^{n}E_{a})(\sum_{b\neq c}^{n}E_{b}E_{c})\nonumber\\
&=&\sum_{a,b\neq c}^{n}E_{a}E_{b}E_{c} \nonumber\\
&=&2\sum_{a,b\neq c}^{n}(\delta_{ab}+c\delta_{ab}E_{a})E_{c}\nonumber\\
&=&2\sum_{a,b\neq c}^{n}\delta_{ab}E_{c}
+2c\sum_{a,b\neq c}^{n}(\delta_{ac}+c\delta_{ac}E_{a})\delta_{ab}\nonumber\\
&=&2\sum_{b\neq c}^{n}\delta_{bb}E_{c}+O(c)=2(n-1)\sum_{c}E_{c},
\eea
so $b_{E_t}=2(n-1)$.

\bea
\tilde{E}_a \Phi&=&(E_{a}-\frac{1}{n}\sum_{b=1}^{n}E_{b})
(\sum_{c\neq d=1}^{n}E_{c}E_{d}) \cr \nonumber\\
&=&\sum_{c\neq a}^{n}E_{a}E_{c}E_{d}
-\frac{1}{n}(\sum_{b=1}^{n}E_{b})(\sum_{c\neq d=1}^{n}E_{c}E_{d}) \cr \nonumber\\
&=&2\sum_{c\neq d}^{n}(\delta_{ac}+c\delta_{ac}E_{a})E_{d}
-\frac{1}{n}(\sum_{b=1}^{n}E_{b})(\sum_{c\neq d=1}^{n}E_{c}E_{d}) \cr \nonumber\\
&=&2\sum_{d\neq a}^{n}E_{d}
-\frac{1}{n}(2n-2)(\sum_{b=1}^{n}E_{b}) + \cdots \cr \nonumber\\
&=&-2(E_{a}-\frac{1}{n}\sum_{b=1}^{n}E_{b}) + \cdots
\eea
so $b_{\tilde{E}_{a}}=-2$. Now we can obtain $x_{E_{t}}$ , $x_{\tilde{E}_{a}}$,

\bea
x_{E_{t}}&=&x_{E_{t}}^{0}+\frac{2b_{E}}{b}y_{g}\\
x_{\tilde{E}_{a}}&=&x_{\tilde{E}_{a}}^{0}+\frac{2b_{\tilde {E}_{a}}}{b}y_{g},
\eea
we have neglected $O(c)$ and $O(c^{2}\cdot\cdot\cdot)$
so $b=2(n-2)$ and in the limit of $n \rightarrow 0$,

\bea
x_{E_t}&=&x_{E}^{0}+(1+ n/2)y_{g}\\
x_{\tilde{E}_{a}}&=&x_{\tilde{E}_{a}}^{0}+ (1-n/2) y_{g},
\eea
Now define  $ x_{E_t} = 2 \Delta_{E_t} $ and $x_{\tilde{E}_{a}}=2 \Delta_{\tilde E_a}$.
\\
The important observation is that the fields
$E_t$ and $\tilde E_a$ have the proper scaling dimensions close to
$n \rightarrow 0$ as $\Delta_{E_t} =
\Delta^{(0)}_{E_a }+ y_g/2 +
O(y_g^2)$
and $\Delta_{\tilde E} =
\Delta^{(0)} _E +  y_g/2 +
O(y_g^2)$ respectively.
It is clear that the singlet field $E_t$ becomes degenerate
with the $(n-1)$ operators
$\tilde E_a$.
However they do not form the basis of the Jordan cell for the dilatation
operator.
 Starting from the following replicated Hamiltonian one can show that
correlation functions calculated against this effective Hamiltonian
will correspond to correlator averaged against the initial Hamiltonian
with quenched disorder,
\bea
&&H_{R} = \sum_{a=1} ^ n H_a + t \int d^d x \sum_{a=1} ^ n E_a(r)
- g \int d^d r \sum_{a\neq b} E_a(r) E_b(r) \cr \nonumber
\\
&& \overline{ < E(r) >_H  } \leftrightarrow \lim_{n \rightarrow 0} < E_a(r) >_{repl}
\cr \nonumber \\
&&\overline{ < E(0) E(r) >_H  } \leftrightarrow \lim_{n \rightarrow 0} < E_a(0) E_a(r) >_{repl}
\cr \nonumber \\
&&\overline{ < E(0) >_H  < E(r) >_H }
 \leftrightarrow \lim_{n \rightarrow 0} < E_a(0) E_b(r) >_{repl}  \hskip 1cm a \neq b
\cr \nonumber \\
&& \overline{< E(r_1) E(r_2) E(r_3) >_H} \leftrightarrow \lim_{n \rightarrow 0}
< E_1 (1) E_1 (2) E_1 (3) >_{repl}
\cr \nonumber \\
&& \overline{< E(1) E(2) >_H < E(3) >_H} \leftrightarrow \lim_{n \rightarrow 0}
  < E_1 (1) E_1 (2) E_2 (3) >_{repl}
\cr \nonumber \\
&& \overline{<E(1)>_H <E(2)>_H <E(3)>_H} \cr \nonumber \\
&& \leftrightarrow \lim_{n \rightarrow 0}
< E_1 (1) E_2 (2) E_3 (3) >_{repl}
\eea
etc.
\\

To find the logarithmic pair consider:

\bea
&&<E_t (0) E_t(r)> = \cr \nonumber \\
&& n ( <E_1(0) E_1(r)> - (n-1) <E_1(0) E_2(r)>
) \cr \nonumber \\
&&\equiv n A(n) r^{-2 \Delta_E (n)} \cr \nonumber \\
&&< \tilde E_a (0) \tilde E_a (r)>
 = \cr \nonumber \\
&& (1- \frac{1}{n})(   <E_1(0) E_1(r)> - <E_1(0) E_2(r)> )
\cr \nonumber \\
\equiv
&&(1-\frac{1}{n}) B(n) r^{-2 \tilde \Delta_E (n)}
\eea

The above equations enable us to write the quenched
averaged connected two-point correlation
functions of energy density operator in terms of
$ <E_1(0) E_1(r)>$ and $ <E_1(0) E_1(r)>$ in the
limit of $n \rightarrow 0$ as:
$\overline{< E(0) E(r) >}_c =  <E_1(0) E_1(r)>  -  <E_1(0) E_2(r)>  $
which is equal $B(0) r^{- 2 \Delta_E}$ and
it has a pure scaling behavior. However the correlation functions
$<E_1(0) E_1(r)>$ and $<E_1(0) E_2(r)>$ have the logarithmic singularities and behave as:
\bea
&& < E_1(0) E_1(r) > =  (A'(0) - B'(0) + B(0)
 \cr \nonumber \\
& -& B(0) \frac{y_g}{2} \ln r)
  r^{-2\Delta_E}  \cr \nonumber \\
&& < E_1(0) E_2(r) > =  (A'(0) - B'(0) \cr \nonumber \\
&&- A(0) \frac{y_g}{2} \ln r)
r^{-2\Delta_E}
\eea
where $A(0) = B(0)$.
The prime sign in the eq. (258) means differentiating with respect to $n$.

 This means that in the limit $ n \rightarrow 0$
the field $E_t$ and $E_a$ form a basis of Jordan cell, i.e. their two point
correlation functions behave as:
$<E_t(0) E_t(r)>=0$, $ <E_t(0) E_a(r) > = a_1 r ^{-2 \Delta_E}$
and $<E_a(0) E_b(r)> = (-2 a_1 \ln r + D_{a,b}) r^{-2 \Delta}$, where
$a_1$ and $D_{a,b}$ are some constants.
\\
 Also the ratio
of quenched averaged two-point correlators of the energy density
operator to  connected one has a universal $r$-dependence as:
\be
\frac{ \overline{ < E(0) E(r) >} }{\overline {< E(0) E(r) >}_c} \sim
\frac{ \overline {< E(0) >< E(r)>}}{ \overline {< E(0) E(r) >}_c} \sim \ln r
\ee
We note that in 2D we have dealt with two-dimensional conformal
 field theory, relying heavily on the underlying Virasoro algebra.
 For an extension to $d$ dimensions one has to modify the
 representation of the Virasoro algebra to higher dimensions.
 We consider a doublet of fields (Jordan cell)

\be
\Phi = \left (\begin{array}{c} E_t \\ E_a
\end{array} \right ),
\ee

 and note that under
 D-dimensional conformal transformation $\bf{ x} \rightarrow \bf{x'}$, we have,
 $\Phi(\bf{x})  \rightarrow \Phi'(\bf{x'}) = G ^T \Phi(\bf{x})$
 where $T$ is a two dimensional matrix which has Jordan form and
  $G=|| \frac{\partial x'}{\partial x}||$  is the Jacobian.
  The fields $E_t$ and $E_a$, transforms as:
\bea
 E_t(\bf{x'}) & = & G^{- \frac{2\Delta_E}{D}} E_t (\bf{x})  \cr \nonumber \\
 E_a(\bf{x'}) & = & G^{- \frac{2\Delta_E}{D}} ( \ln (G) E_t (x) + E_a (x) )
\eea
This expresses that the top-field $E_t$ always transforms as an ordinary
scaling operator. It can be verified that the correlation functions of
fields $E_t$ and $E_a$ have the standard
d- dimensional logarithmic conformal field theory structures.

Let us consider the case that $d=2$. Therefore we can write:
\bea
L_0 E_t &=& \Delta_E E_t \cr \nonumber
L_0 E_a &=& \Delta_E E_a + E_t \cr \nonumber
L_0 E_b &=& \Delta_E E_b + E_t \cr \nonumber
L_0 E_c &=& \Delta_E E_c + E_t
\eea
where we have used the replica symmetry.
 Using the above equations, it is evident that the dimension
of difference-fields $ E_a - E_b $ with $a \neq b$ is $\Delta_E$ and it
transforms as an ordinary operator under the scaling transformation.
The important observation is that the individual logarithmic operator $E_a$
do not contribute to the $connected$ quenched averaged
correlation functions.
Instead the connected averaged correlation functions depend on the
difference fields $ E_a - E_b $ only.
For instance in the following we write
the
connected quenched averaged 2,3 and 4-point functions of local energy
density in terms of the
difference operators explicitly,
\bea
\overline {< E(1) E(2) >}_c &=& \frac{1}{2} < (E_a-E_b)_{(1)} (E_a -E_b)_{(2)}>
\eea
\bea
\overline{< E(1) E(2) E(3) >}_c& = &
< (E_a-E_b)_{(1)} \cr \nonumber \\
&& (E_a-E_c)_{(2)} (E_a-E_b)_{(3)}>
\eea
\bea
&& \overline{< E(1) E(2) E(3) E(4)>}_c = \cr \nonumber \\
&&< (E_a-E_b)_{(1)} (E_a-E_c)_{(2)}  (E_a-E_d)_{(3)} (E_a-E_b)_{(4)} > \cr \nonumber \\
&&-\frac{1}{2} < (E_a-E_b)_{(1)} (E_c-E_d)_{(2)}  (E_c-E_d)_{(3)} (E_a-E_b)_{(4)} > \cr \nonumber \\
&&-\frac{1}{4} < (E_a-E_b)_{(1)} (E_c-E_d)_{(2)}  (E_a-E_b)_{(3)} (E_c-E_d)_{(4)} > \cr \nonumber \\
&&-\frac{1}{4} < (E_a-E_b)_{(1)} (E_a-E_b)_{(2)}
 (E_c-E_d)_{(3)} (E_c-E_d)_{(4)} >  \nonumber
\eea where the last equation has only 15 independent terms. This
shows that quenched averaged connected correlation functions have
a pure scaling behavior. Let us verify this results from direct
calculation of quenched averaged connected 3-point correlation
function of energy density.

We are interested in exact derivation of the various 3-point quenched
averaged functions as $ \overline {< E(1) E(2) E(3) >}$,
$ \overline{<E(1) E(2)><E(3)>}$ and
$\overline{<E(1)><E(2)><E(3)>}$, which can be written in terms of the replica
correlation functions $ < E_1 (1) E_1 (2) E_1 (3) > = a $
$ < E_1 (1) E_1 (2) E_2 (3) > = b $ and
$ < E_1 (1) E_2 (2) E_3 (3) > = c $, respectively.
One can derive the correlation functions $a$, $b$ and $c$ by means
of the 3-point functions of $E_t$ and $\tilde E_a$ as:
\bea
&&< E_t(1) E_t(2) E_t(3) > = na+3n(n-1)b \cr \nonumber \\
&&n(n-1)(n-2) c \equiv n A_1
\eea
\bea
&&< \tilde E_a(1) \tilde E_a(2) E_t(3) > = n_1 a +( n_1 ^2 (n-1) \cr \nonumber \\
&& -4n_1 ^2 + \frac{1}{n^2} (n-1)^2 + \frac{2}{n^2} (n-1)(n-2))b
\cr \nonumber \\
&+& ( -\frac{2}{n}
n_1 (n-1) (n-2) + \frac{1}{n^2} (n-2)^2 (n-1) ) c
\cr \nonumber \\ &&\equiv (1- \frac{1}{n}) B_1
\eea
and finally,
\bea
&&< \tilde E_a(1) \tilde E_a(2) \tilde E_a(3) > = (n_1^2 - \frac{n-1}{n^3}) a
 \cr \nonumber \\
&& (-3 n_1^2 \frac{n-1}{n} -\frac{3}{n^3} (n-1)(n-2) + \frac{3}{n^2} n_1 (n-1))b
  \cr \nonumber \\
&& + ( \frac{3}{n^2} n_1 (n-1) (n-2) - \frac{1}{n^3} (n-1)(n-2)(n-3) ) c
 \cr \nonumber \\
 &&\equiv (1- \frac{1}{n}) (1- \frac{2}{n}) C_1
\eea
where $n_1 = (1 - \frac{1}{n})$.
To derive the above equations we use the replica symmetry and
symmetries of the
various type of 3-point correlation functions under interchanging of
positions.
We note that replica symmetry leads to
have $ <\tilde E_a(1) E_t(2) E_t(3) > = 0 $ and therefore,
dose not give any new relationship
between $a$, $b$ and $c$.
Using the above equations, it can be found that
the correlation functions $a$, $b$ and $c$ are as following:
\bea
a &=& \frac{3nB_1 -3nC_1 + n^2 C_1 +A_1 - 3B_1 +2 C_1}{n^2} \cr \nonumber \\
b &=& \frac{nB_1 - n C_1 +A_1 - 3B_1 +2 C_1}{n^2} \cr \nonumber \\
c &=& \frac{A_1 - 3B_1 +2 C_1}{n^2} \eea Where $A_1$ , $B_1$ and
$C_1$ are pure scaling functions of variables $r_{i,j}$. Using the
above equations we can show that the connected quenched averaged
3-point function behaves as: \be \overline{< E(1) E(2) E(3) >_c} =
2 c + a -3b = C_1 \ee which is a scaling function and confirms the
observation that the logarithmic operators (individually) do not
play any role in the connected quenched averaged correlation
functions. In addition one can derive the correlation functions $<
E_i (1) E_j (2) E_k (3)> $ for given $i$, $j$ and $k$ in the limit
of $n \rightarrow 0$ and show that they have the following form:
\bea
<E_i(1) E_j(2) E_k(3)> & = & [ \alpha_{ijk} - \beta_{ijk} D_1 \cr \nonumber \\
&+& \gamma_{ijk} (4D_2 - D_1 ^2)] f(1,2,3)
\eea
where $f(1,2,3) = ( r_{12} r_{13} r_{23} ) ^{-2 \Delta_E}$
 , $D_1 = \ln(r_{12} r_{13} r_{23})$ and
 $D_2 = \ln r_{23} \ln r_{13} + \ln r_{13} \ln r_{12} + \ln r_{23} \ln r_{12} $.
It can also be shown that the ratio of various symmetrized 3-point
functions to the
connected one behaves asymptotically as a $universal$ function
\be
\frac{1}{3}(4 D_2 - D_1^2).
\ee
We generalize the above calculations to derive the
various type of 4-point correlation functions and show that
the ratio of the various disconnected to the connected one
has the following universal asymptotic:
\be
\sim \frac {1}{36} [ O_1 ^3 - 6 O_2 - 3 O_3 - 12 O_4 - 18 O_5]
\ee
where $O_1= \ln ( r_{12} r_{13} r_{14} r_{23} r_{24} r_{34})$,
$O_2= (\ln r_{ij} \ln r_{kl} ^2 + \cdots)$ with $ i\neq j \neq k \neq l$,
$O_3 = (\ln r_{ij} \ln r_{ik} ^2 + \cdots)$ with $ i\neq j \neq k$,
$O_4 = (\ln r_{ij} \ln r_{kl} \ln r_{lj} + \cdots)$ with $ i\neq j \neq k \neq l$,
and finally
$O_5 = (\ln r_{ij} \ln r_{ik} \ln r_{il} + \cdots)$ with $ i\neq j \neq k \neq l$

One can check directly that these different types of the 3 and
4-point correlation functions have the general property of a
logarithmic conformal field theory that the logarithmic partner
can be regarded as the formal derivative of the ordinary fields
(top field) with respect to their conformal weight. In this case,
one can consider the $E_a$ fields as the derivative of $E_t$ with
respect to $n$ . We emphasis that the derivative with respect to
scaling weight can be written in terms of the derivative with
respect to $n$. These properties enable us to calculate any
N-point correlation function containing the logarithmic field
$E_a$ in terms of the correlation functions of the top-fields. We
have shown that the individual logarithmic operators $E_a$ do not
have any contribution to the quenched averaged connected
correlation functions of the energy density. We also obtain that
the connected correlation functions can be written in terms of the
difference fields which transform as an ordinary scaling operator.
However they will play a crucial role to the disconnected averaged
correlation functions. Also we find that the ratio of the various
types of 3 and 4-point quenched averaged correlation functions to
the connected ones have a universal asymptotic behavior and give
their explicit form.
Our analysis are valid in all dimensions as long as the dimension
is below the upper critical dimensions. To derive the above
results we have used the replica symmetry. Any attempt towards the
breaking of this symmetry will change the above picture and may
produces more than one logarithmic fields in the block and produce
higher order logarithmic singularities. These results can be
easily generalized to other problem such as polymer statistics,
percolation and random phase sine-Gordon model etc.

{\bf Acknowledgement}\\

 We thank
N. Abedpour, A. Aghamohammadi, F. Azami, V. Karimipour,
 F. Kheirandish,
 A.A. Masoudi, S. Rouhani and  F. Shahbazi
for useful discussions and comments.


\newpage

\begin{thebibliography}{99}
\bibitem{1} John Cardy " Scaling and Renormalization in Statistical Physics"
Cambridge uni. press, (1996)
\bibitem{2} A.B. Harris, J.Phys. {\bf C7} (1974) 1671
\bibitem{3} B.N. Shalaev, Phys. Reports {\bf 273}(3) (1994) 129
\bibitem{4} D. Bernard, hep-th/9509137
\bibitem{5} John Cardy " Logarithmic Correlaions in Quenched Random Magnets
and Polymers" cond-mat/9911024; cond-mat/0111031
\bibitem{6} V. Gurarie, Nucl. Phys. {\bf B410} (1993) 535
\bibitem{7} M.R. Rahimi Tabar, A. Aghamohammadi and M. Khorrami
 Nucl. Phys. {\bf B497} (1997) 555
\bibitem{8} P. di Francesco, etal. {\it Conformal Field Theory}, Springer Verlag
(1997)
\bibitem{9} C. Itzykson and J.M. Drouffe, {\it Statistical Field Theory}
Cambridge (1998)
\bibitem{10} M. Henkel, {\it Conformal Invariance and Critical Phenomena}
, Springer (1999)
\bibitem{11} D. Bernard, Z. Maassarani and P. Mathieu,
 Mod.Phys.Lett. {\bf A12} 535 (1997)
\bibitem{12}  I.I. Kogan, A. Lewis and O.A. Soloviev, Int.J.Mod.Phys {\bf A13}
 1345 (1998)
\bibitem{13} A. Nichols and S. Sanjay, Nucl.Phys.{\bf B597} 633 (2001)
\bibitem{14} A. Nichols, hep-th/0007007
\bibitem{15} M.R. Gaberdiel, hep-th/0105046
\bibitem{16} Rozansky and H. Saleur, Nucl.Phys.{\bf B376} 461 (1992);
Nucl.Phys.{\bf B389} 365 (1993)
\bibitem{17} J.S. Caux, I.I. Kogan, A. Lewis and A.M. Tsvelik,
Nucl.Phys.{\bf B489} 469 (1997)
\bibitem{18} M. Khorrami, A. Aghamohammadi and A.M. Ghezelbash,
 Phys.Lett.{\bf B 439} 283 (1998)
\bibitem{19} F. Kheirandish and M. Khorrami, Eur.Phys.J.{\bf C18} 795 (2001);
Eur.Phys.J.{\bf C20} 593 (2001)
\bibitem{20} A.W.W. Ludwig, cond-mat/0012189
\bibitem{21} M.J. Bhaseen,
Nucl.Phys.{\bf B604} 537 (2001)
\bibitem{22} I.I. Kogan and A. Nichols, hep-th/0107160
\bibitem{23} M. Flohr, Mod.Phys.Lett.{\bf A 11} 55 (1996)
\bibitem{24} K. Ino, Phys.Rev.Lett.{\bf 81} 1078 (1998);
Phys.Rev.Lett.{\bf 82} 4902 (1999);
Phys.Rev.Lett.{\bf 83} 3526 (1999);
Phys.Rev.Lett.{\bf 84} 201 (2000);
Phys.Rev.Lett.{\bf 86} 882 (2001)
\bibitem{25} I.I. Kogan and A.M. Tsvelik, Mod.Phys.Lett.{\bf A15} 931 (2000)
\bibitem{26} V. Gurarie, M. Flohr and C. Nayak,
Nucl.Phys.{\bf B498} 513 (1997)
\bibitem{27} A. Cappelli, L.S. Georgiev and I.T. Todorov, Comm.Math.Phys.{\bf 205}
657 (1999)
\bibitem{28} J.S. Caux, Phys.Rev.Lett.{\bf 81} 4196 (1998)
\bibitem{29} M.R. Rahimi Tabar and S. Rouhani, Nouvo Cimento {\bf B 112} 1079 (1997)
Eur.Phys.Lett.{\bf C 37} 477 (1997); Phys.Lett. {\bf A 224} 331 (1997)
\bibitem{30} M. Flohr,
Nucl.Phys.{\bf B482} 567 (1996)
\bibitem{31} S. Skoulakis and S. Thomas, Phys.Lett.{\bf 438} 301 (1998)
\bibitem{32} A. Bilal and I.I. Kogan,
Nucl.Phys.{\bf B449} 569 (1995)
\bibitem{33} H. Saleur,
Nucl.Phys.{\bf B489} 486 (1992)
\bibitem{34} E.V. Ivashkevich, J.Phys.{\bf A32} 1691 (1999)
\bibitem{35} S. Mathieu and P. Ruelle, hep-th/0107150
\bibitem{36} I.I. Kogan and N.E. Movromatos, Phys.Lett.{\bf B375} 11 (1996)
\bibitem{37} J.R. Ellis, N.E. Movromatos and D.V. Nanopoulos,
Int.J.Mod.Phys.{\bf A12} 2639 (1997);
Int.J.Mod.Phys.{\bf A13} 1059 (1998);
Gen.Rel.Grav.{\bf 32} 943 (2000);
Gen.Rel.Grav.{\bf 32} 1777 (2000);
Phys.Rev. {\bf D62} 084019 (2000)
\bibitem{38} I.I. Kogan, N.E. Movromatos and J.F. Wheater,
Phys.Lett.{\bf B387} 483 (1996)
\bibitem{39} J.R. Ellis, N.E. Movromatos and E. Winstanley,
Phys.Lett.{\bf B476} 165 (2000)
\bibitem{40}  N.E. Movromatos and R.J. Szabo, Phys.Rev.{\bf D 59} 104018 (1999);
hep-th/0106259
\bibitem{41} G.K. Leontaris and N.E. Movromatos,
Phys.Rev.{\bf D 61} 124004 (2000);
Phys.Rev.{\bf D 64} 024008 (2001)
\bibitem{42} A. Campbell-Smith and N.E. Movromatos,
Phys.Lett.{\bf B476} 149 (2000);
Phys.Lett.{\bf B488} 199 (2000)
\bibitem{43}  A. Lewis,
Phys.Lett.{\bf B480} 348 (2000)
\bibitem{44} E. Gravanis and N.E. Movromatos, hep-th/0106146
\bibitem{45}
A.M. Ghezelbash, M. Khorrami and A. Aghamohammadi, Int.J.Mod.Phys.{\bf A14}
2581 (1999)
 \bibitem{46} J.Y. Kim, H.W. Lee and Y.S. Myung, hep-th/9812016
\bibitem{47} K. Kaviani and A.M. Ghezelbash, Phys.Lett.{\bf B469} 81 (1999)
\bibitem{48} I.I. Kogan, Phys.Lett.{\bf B456} 66 (1999)
\bibitem{49} Y.S. Myung and H.W. Lee, JHEP. {\bf 9910} 009 (1999)
\bibitem{50} I.I. Kogan and D. Polyakov, Int.J.Mod.Phys.{\bf A16} 2559 (2001)
\bibitem{51} S. Sanjay, hep-th/0011056
\bibitem{52} S. Moghimi-Araghi, S. Rouhani and M. Saadat,
hep-th/0105123
\bibitem{53} M. Flohr Phys.Lett.{\bf B444} 179 (1998)
\bibitem{54} J.S. Caux, I.I. Kogan, A.M. Tsvelik, Phys.Lett.{\bf B466} 444 (1996)
\bibitem{55} I.I. Kogan, C. Mudry, A.M. Tsvelik,
Phys.Rev.Lett.{\bf 77} 707 (1996)
\bibitem{56}   Z. Maassarani and D. Serban, Nucl.Phys.{\bf B489} 603 (1997)
\bibitem{57}   V. Gurarie, Nucl.Phys.{\bf B546} 765 (1999)
\bibitem{58}   J.S. Caux, N. Taniguchi and A.M. Tsvelik, Nucl.Phys.{\bf B525} 671
 (1998)
\bibitem{59} T. Davis and J. Cardy, Nucl. Phys. {\bf B 570} 713 (2000)
\bibitem{60} M. Reza Rahimi Tabar, Nucl. Phys. {\bf B588} 630 (2000)
\bibitem{61}   V. Gurarie and A. Ludwig, cond-mat/9911392
\bibitem{62} M.J. Bhaseen, J.S. Caux, I.I. Kogan and
A.M. Tsvelik, cond-mat/0012240
\bibitem{63} M.J. Bhaseen, cond-mat/0012420
\bibitem{64} M. Flohr, Mod.Phys.Lett,{\bf A11} 4147 (1996); Int.J.Mod.Phys.
{\bf A12} 1943 (1997); Nucl.Phys. {\bf B514} 523 (1998); hep-th/0009137;
hep-th/0107242.
\bibitem{65} H.G. Kausch, hep-th/9510149; Nucl.Phys. {\bf B583} 513 (2000)
\bibitem{66} M.R. Gaberdiel and H.G. Kausch,  Nucl.Phys. {\bf B477} 293 (1996);
Phys. Lett. {\bf B386} 131 (1996); Nucl.Phys. {\bf B538} 613 (1999)
\bibitem{67} F. Rohsiepe, hep-th/9611160
\bibitem{68} A.M. Ghezelbash and V. Karimipour, Phys.Lett.{\bf B402} 282 (1997)
\bibitem{69} I.I. Kogan and A. Lewis, Nucl.Phys. {\bf B509} 687 (1998);
Phys.Lett. {\bf B431} 77 (1998)
\bibitem{70} M.R. Rahimi Tabar and S. Rouhani, Phys.Lett. {\bf B431} 85 (1998)
\bibitem{71} W. Eholzer, L. Feher and A. Honecker, Nucl.Phys. {\bf B518} 669 (1998)
\bibitem{72} M. Khorrami, A. Aghamohammadi and M.R. Rahimi Tabar,
Phys.Lett. {\bf B419} 179 (1998)
\bibitem{73} A. Aghamohammadi, M. Alimohammadi and M. khorrami,
Mod.Phys.Lett. {\bf A12} 1349 (1997)
\bibitem{74} N.E. Mavromatos and R.J. Szabo,
Phys.Lett. {\bf B430} 94 (1998)
\bibitem{75} A. Shafiekhani and M.R. Rahimi Tabar, Int.J.Mod. Phys. {\bf A12} 3723
(1997)
\bibitem{76} S. Moghimi-Araghi and S. Rouhani, Lett.Math.Phys. {\bf 53} 49 (2000)
\bibitem{77} G. Giribet, Mod.Phys.Lett. {\bf A16} 821 (2001)
\bibitem{78}
S. Moghimi-Araghi, S. Rouhani, Lett.Math.Phys.{\bf 53} 49 (2000)
\bibitem{79} I.I. Kogan and J.F. Wheater, Phys.Lett.{\bf B486} 353 (2000)
\bibitem{80} A. Lewis, hep-th/0009096
\bibitem{81} Y. Ishimoto, hep-th/0103064
\bibitem{82} S. Kawai and J.F. Wheater, Phys.Lett.{\bf B508} 203 (2001)
\bibitem{83} H. Hata and S. Yamaguchi, hep-th/0004189
\bibitem{84} M. Flohr,  hep-th/0107242
\end{thebibliography}
\end{document}